\documentclass[aps,prd,twoside,showpacs,superscriptaddress,twocolumn,nofootinbib,floatfix,preprintnumbers]{revtex4}
\usepackage{graphics}
\usepackage{graphicx}
\usepackage{epsfig}
\usepackage{psfrag}

\usepackage{amsfonts}
\usepackage{amssymb}

\newcommand{\eref}[1]{(\ref{#1})}

\newcommand{\comment}[1]{}

\preprint{DFPD 06/TH 07}
\preprint{CPHT-RR022.0506}

\begin{document}
\title{Lepton Flavour Violation from SUSY--GUTs:\\
 Where do we stand for MEG, PRISM/PRIME and a Super Flavour factory}
\author{L. Calibbi}
\affiliation{ Dipartimento di Fisica `G. Galilei' and
INFN, Sezione di Padova,
Universit\`a di Padova, Via Marzolo 8, I-35131, Padova, Italy.}
\author{A. Faccia}
\affiliation{ Dipartimento di Fisica `G. Galilei' and
INFN, Sezione di Padova,
Universit\`a di Padova, Via Marzolo 8, I-35131, Padova, Italy.}
\author{A. Masiero}
\affiliation{ Dipartimento di Fisica `G. Galilei' and
INFN, Sezione di Padova,
Universit\`a di Padova, Via Marzolo 8, I-35131, Padova, Italy.}
\author{S. K. Vempati}
\affiliation{Centre de Physique Theorique
\footnote{Unit{\'e} mixte du CNRS et de l'EP, UMR 7644.},
Ecole Polytechnique-CPHT, 91128 Palaiseau Cedex, France}
\affiliation{The Institute of Mathematical Sciences, Chennai 600 013, India}

\begin{abstract}
We analyse the complementarity between  Lepton Flavour Violation (LFV) and 
LHC experiments in probing the Supersymmetric (SUSY) Grand Unified Theories 
(GUT) when neutrinos got a mass via the see--saw mechanism. Our analysis is
performed in an $SO(10)$ framework, where at least one neutrino 
Yukawa coupling
is necessarily as large as the top Yukawa coupling. 
Our study thoroughly takes into account the whole RG
running, including the GUT and the right handed neutrino
mass scales, as well as the running of the observable
neutrino spectrum.
We find that the upcoming
(MEG, SuperKEKB) and future (PRISM/PRIME, Super Flavour factory) LFV 
experiments will be able to test such SUSY framework for SUSY masses to be 
explored at the LHC and, in some cases, even beyond the LHC sensitivity reach.
\end{abstract}


\maketitle

\section{Introduction}

In this paper we address the issue of detecting low--energy supersymmetry
(SUSY)
at present, upcoming and planned Lepton Flavour Violation (LFV)
experiments. Moreover we evaluate the complementarity between 
these experiments and the CERN Large Hadron Collider (LHC) experiments
as probes of supersymmetric grand--unified (SUSY--GUT) scenarios.

The study of Flavour Changing Neutral Current (FCNC) processes, which are 
suppressed by the Glashow--Iliopolous--Maiani  mechanism \cite{GIM}
in the Standard Model (SM) of particle physics, 
has been considered 
for a long time  a powerful tool in order to shed light on new physics,
especially for testing low--energy supersymmetry. Indeed, 
taking into account the fact that neutrinos have mass and mix
\cite{sol, SKsolaratm, SNO123, KamLAND, K2K},
the Standard Model predicts Lepton Flavour Violating  processes
in the charged sector to occur at a negligible rate 
(e.g. $BR(\mu\to e\,\gamma)\sim\mathcal{O}(10^{-54})$
\cite{petkov, chengli}). Given the future experimental
sensitivities to LFV processes (Table \ref{lfvtable}), 
the discovery of such processes
will open a window to new physics.

\begin{table}[b]
\caption{\label{lfvtable}Present bounds and expected experimental sensitivities
 on LFV processes
\cite{mega, sindrum2, belletmg, belle2, babar, belletalk,%
meg, superKEKB, prismprime:loi, prismprime:kuno, superflavour}.} 
\begin{ruledtabular}
\begin{tabular}{lcc}
Process & Present bound & Future sensitivity  \\ 
\hline
  BR($\mu \to e\,\gamma$) & $1.2~ \times~ 10^{-11}$  &
$\mathcal{O}(10^{-13} - 10^{-14})$ \\ 
  BR($\mu \to e\,e\,e$ ) & $1.1~ \times~ 10^{-12}$ &
$\mathcal{O}(10^{-13} - 10^{-14})$ \\ 
  CR($\mu \to e$ in Ti) & $4.3~ \times~ 10^{-12}$ &
$\mathcal{O}(10^{-18})$\footnotemark[1] \\ 
  BR($\tau \to e\,\gamma$) & $3.1~ \times~ 10^{-7}$ &
$\mathcal{O}(10^{-8})-\mathcal{O}(10^{-9})$\footnotemark[1]   \\ 
  BR($\tau \to e\,e\,e$) & $2.7~ \times~ 10^{-7}$ &
$\mathcal{O}(10^{-8})-\mathcal{O}(10^{-9})$\footnotemark[1]   \\ 
  BR($\tau \to \mu\,\gamma$) & $6.8~ \times~ 10^{-8}$ &
$\mathcal{O}(10^{-8}) -\mathcal{O}(10^{-9})$\footnotemark[1]  \\ 
  BR($\tau \to \mu\, \mu\, \mu$) & $2~ \times~ 10^{-7}$ &
$\mathcal{O}(10^{-8}) -\mathcal{O}(10^{-9})$\footnotemark[1]   \\ 
\end{tabular}
\end{ruledtabular}
\footnotetext[1]{Planned or discussed experiment, not yet under construction}
\end{table}

It is known that a generic low--energy SUSY model (i.e. a model with arbitrary 
mixings in the 
soft breaking parameters sector) would induce unacceptably large flavour 
violating effects \cite{gabbianimasiero96}.
The  unobserved departures from SM in FCNCs makes it reasonable to assume
 flavour 
universality in the mechanism that breaks SUSY. 
On the other hand, even taking flavour universal 
SUSY breaking boundary conditions, renormalization
effects can generate sizable flavour mixings 
in the running of the soft parameters from the SUSY breaking mediation scale
down to the SUSY decoupling scales.
 In the leptonic sector, the relevance
of such effects strongly depends on the
neutrinos' parameters.

The existence and smallness
of neutrinos' masses can be simply explained by the see--saw mechanism
\cite{seesawpapers}, by
introducing Right--handed  Neutrino (RN) fields, that are singlets under the SM
gauge transformations. Since there is no gauge symmetry that
protects them, the RNs can get a large Majorana mass 
$(\hat{M}_R)_{ij}$, breaking
the conservation of lepton number. When they are integrated out, they will give
rise to an effective light neutrino Majorana mass matrix
\begin{equation}
\label{see-saw}
m_\nu = - Y_\nu \hat{M}^{-1}_R Y_\nu^T \langle H_u \rangle^2 \:,
\end{equation}
where $(Y_{\nu})_{ij}$ are the Yukawa couplings between left and
right handed neutrinos and  $\langle H_u \rangle$ is 
the Vacuum Expectation Value (VEV) acquired by the up sector  Higgs 
field.

It is known \cite{borzumatimasiero}
 that the marriage between see--saw and SUSY can
generate observable LFV rates in the charged lepton sector.
In their renormalization group (RG) evolution, the slepton soft
masses $(m^2_{\tilde{L}})_{ij}$ acquire LFV entries that are proportional to 
$(Y_\nu Y_\nu^\dagger)$
\begin{equation}
\label{mLfromseesaw}
(m^2_{\tilde{L}})_{i\neq j} =
- \frac{3m^2_0 + A^2_0}{16\pi^2} \sum_k Y_{\nu \: ik} 
Y_{\nu \: kj}^\dagger
\ln \left(\frac{M^2_X}{M^2_{R \: k}} \right) 
\end{equation}
where  $M_{R_k}$ is the mass of the $k$-th right handed neutrino
($i,~j,~k$ being generation indices), $m_0$ and
$A_0$ are the universal supersymmetry breaking scalar masses and
scalar trilinear couplings respectively.
Since the see--saw equation
\eref{see-saw} allows large neutrinos' Yukawa couplings,
 sizable effects can stem from this running.
From the above it is obvious that any estimate of 
$(m^2_{\tilde{L}})_{i\neq j}$
would require a complete knowledge of the neutrino Yukawa matrix
$(Y_\nu)_{ij}$ which is not fixed by the see--saw equation, even with an
improved knowledge of the neutrino oscillation parameters,
as in \eref{see-saw} there is a mismatch between the number
of unknowns and that of low energy observables. 
This could indeed pose a problem compared to the quark--squark
FCNCs sector, where it is possible to make firm predictions
of FV entries due to RG evolution in a  flavour universal
boundary condition.
On the other hand, in the quark sector
the disentangling of FV effects stemming from SUSY from
those coming from the SM is
more problematic, as both are driven by the CKM mixing
matrix and happen to be roughly of the same order of
magnitude. 
In the charged lepton sector we have the opposite
scenario: SM contributions are well below any envisageable experimental
sensitivity, but it is not possible to predict SUSY induced
FCNC rates without
resorting to an ansatz with regard to the form of the
Yukawa matrix $Y_\nu$ or the general framework of the theory.
Many groups \cite{many} have addressed this issue in different
frameworks.

In this paper we inspect LFV in a $SO(10)$ supersymmetric
grand unified (SUSY--GUT) framework. The choice of
the GUT scenario stems from the fact that the possible
detection of SUSY particles at the LHC will provide an
indirect evidence for such a scenario. Moreover in $SO(10)$ theories 
the see--saw mechanism is naturally present and the neutrino
Yukawa couplings are related to those of the up quarks, 
making them naturally large \cite{masivemp02}, so that
sizable LFV entries will stem from the \eref{mLfromseesaw}
RG evolution.
Even if the  $SO(10)$ framework gives us some hints about the unknown neutrino 
Yukawa matrix $Y_{\nu}$, telling us that the eigenvalues are related to the 
ones of the up Yukawa matrix $Y_u$, it 
still leaves uncertainty about the size of mixing 
angles in $Y_\nu$,
 as the knowledge of the low--energy neutrino parameters
(masses and mixings) is not sufficient to set the
 matrices that diagonalize $Y_{\nu}$.  
Following the scheme of previous works
\cite{masivemp02, masivemp04bis, masivemp04}, we bypass
the ignorance about the mixings by considering two extremal benchmark cases.
Such cases are intended as
boundary conditions at high scale.
As a minimal mixing case we take the one in which the neutrino and the
up Yukawa unify at the high scale, so that the mixing is given by the
CKM matrix; this case is named `CKM--case'. As a maximal mixing scenario
we take the one in which the observed 
neutrino mixing is coming entirely from the neutrino Yukawa matrix, so that 
$Y_\nu = U_{PMNS} \cdot Y^{\mathrm{diag}}_u$, where $U_{PMNS}$ is the
Pontecorvo--Maki--Nakagawa--Sakata matrix; in this case the 
   unknown $U_{e3}$ PMNS matrix element turns out to be crucial in
evaluating the size of LFV effects. The maximal case is named
`PMNS--case'.

The aim of the present work is to study, in a minimal supergravity
(mSUGRA)
scenario \footnote{We are taking the soft trilinear mass scale
$A_0$ as a free parameter, not linked
to the Higgs sector $B$ parameter as in a strict mSUGRA scenario}, 
the influence on LFV from the RG
running both above and below the unification scale $M_{GUT}$.
A preliminary version of this analysis is already present in the literature
\cite{masivemp04}. However that work considered the LFV contributions
from the see--saw structure only. It is well known that in a 
$SO(10)$ scenario there are other contribution stemming from
the grand unified structure \cite{barbierihall}, as on general grounds the 
SUSY breaking mediation scale  and the GUT scale do not coincide. 
In the present work we detail the various contributions and their relative
relevance as source of LFV in the $SO(10)$ framework.
We  analyse in detail
the impact of LFV experiments in the parameter space region
 that will be probed by the LHC,
both in the minimal (CKM--like) and in the maximal (PMNS--like)
cases and for different values of $\tan\beta$.
Such a \textit{vis--a--vis} analysis allows us to
address the issue of the complementarity between LHC and
LFV experiments as probes of SUSY--GUTs.

We argue that, even in presence of a discovery machine like 
the LHC, flavour physics experiments will still play an important role in the
hunt for new physics. 
Indeed, LHC evidences alone will hardly discriminate among 
the many possible SUSY realizations, while flavour physics should be, 
in this sense, more sensitive. Moreover, several flavour physics 
experiments are currently running or under construction 
(such as B--factories, the SuperKEKB upgrade \cite{superKEKB} 
and
 the upcoming MEG \cite{meg} experiment at PSI), 
and thus some hints of new physics before the LHC era 
are also possible. Furthermore, there are discussions 
and plans on very sensitive LFV experiments 
beyond the LHC era, such as
the PRISM/PRIME experiment at J--PARC \cite{prismprime:loi}
and a Super Flavour factory \cite{superflavour}. It 
is thus timely to ask what will be the capability of
such experiments to discriminate between different
SUSY--GUT realizations, in the case that the LHC gets a positive
evidence for SUSY.
Let us note that, even in the case that nothing is seen at the LHC, 
taking into account that SUSY effects decouple slowly with increasing
SUSY masses, flavour physics could still exhibit some indirect SUSY evidence.

The main results of our analysis are:

\begin{itemize}
\item The maximal PMNS case is already ruled out by the current
MEGA bound on $\mu \to e \, \gamma$ in the case the 
squark masses are lighter than 1.5 TeV. MEG will improve the situation
by testing it well beyond the reach of LHC sensitivity.

\item If the unknown $U_{e3}$ angle is very small, at present
the PMNS case is constrained only in the high $\tan\beta$ region, 
by B--factories BR($\tau \to \mu \, \gamma$) bounds, to lay in the
region of squark masses bigger than  800 GeV. In the future, MEG
will be able to test this scenaro in all the LHC accesible
SUSY parameter space if $\tan\beta$ is high. For small $\tan\beta$
the best probe will come from SuperKEKB $\tau \to \mu
\, \gamma$ BR bounds, testing it for squark masses up
to 700 GeV.

\item The minimal CKM case is at present unconstrained. MEG
will be able to test it in the high $\tan\beta$ region and
for squark masses lighter than 800 GeV; this scenario will
evade detection by SuperKEKB. The low $\tan\beta$ minimal
mixing case will remain unconstrained.

\item The proposed post--LHC era PRISM/PRIME and
Super Flavour factory experiments will much improve
the situation. PRISM/PRIME would supersede MEG by
testing, by mean of $\mu \to e$ conversion in Ti,
all the scenarios in all the LHC accessible SUSY
parameter space. A Super Flavour factory would be
higly complementary, being able to detect the
LFV $\tau \to \mu \, \gamma$ process up to 1 TeV
squark masses.
\end{itemize}

The paper is organized as follows: in section II we
motivate, in the context of SUSY--GUTs, our $Y_\nu \sim Y_u$ ansatz;
 in section III
we proceed to estimate the LFV RG induced soft masses and the 
branching ratios that stem from them. The numerical routine is
presented in section IV and the results are discussed in section V:
we evaluate LFV rates for the parameter space region within
the reach of the LHC  and comment on the complementarity
between direct SUSY searches and LFV experiments.
In section VI we predict LFV rates at the SPS benchmark points
in our SUSY--GUT frameworks, and argue on the possibility
of future experiments to test these scenarios.
In section VII we give a summary of our findings and draw the conclusions.
Last, in the appendix we fix the notation and give the 
$SU(5)_{\mathrm{RN}}$
RG equations.

\section{See--saw and SUSY $SO(10)$}

An $SO(10)$ SUSY-GUT framework naturally incorporates 
the see--saw mechanism. This is
because the matter representation is a $16$-dimensional spinor
containing  right handed neutrinos which are absent by
choice in the Standard Model spectrum. Further models of $SO(10)$
have two salient features which make them interesting to study
SUSY see--saw: 

(i) Firstly, they unify the Dirac neutrino Yukawa
couplings ($Y_\nu$ ) and the up-type Yukawa couplings ($Y_u$).
Though this unification is exact for smaller Higgs representations,
like the \textbf{10}s, it can be shown that even in the presence of
larger representations like \textbf{120} or \textbf{126},  \textit{at least}
one of the neutrino Yukawa couplings is as large as the top Yukawa.
This can be shown by a simple analysis of the resultant mass matrices
\cite{masivemp02}.

(ii). Secondly, as mentioned in the introduction, unlike the quark
sector, the leptonic sector has the see--saw mechanism that
makes it distinct. Particulary, the observed large neutrino mixing
 doesn't necessarily
mean a large `left' mixing to be present in the neutrino Dirac Yukawa
couplings: in fact 
even CKM--like small mixings in the neutrino Yukawa
couplings can lead to large neutrino mixing \cite{altff}.
This is best depicted by
the Casas--Ibarra parametrization \cite{casasibarrapaper} that solves the
see--saw equation \eref{see-saw} for the neutrino Yukawa
matrix $Y_\nu$
\begin{equation}
Y_\nu = \frac{1}{\langle H_u \rangle} 
U_{PMNS} \mathcal{D}_\nu R \mathcal{D}_N
\label{casas-ibarra}
\end{equation}
where $\mathcal{D}_N$ and $\mathcal{D}_\nu$ are the square root of
the diagonal right handed Majorana and the low energy 
neutrino mass matrices respectively.
 The unknown
complex orthogonal matrix $R$ parametrizes the uncertainty of
the mixing between Majorana and Dirac right handed eigenstates.
This means that if $R$ is the identity the Dirac neutrino Yukawa
matrix inherits the PMNS mixing structure, whereas 
small CKM--like $Y_\nu$ mixings reflect in a non trivial
structure of the misalignment matrix $R$. With this in mind,
most flavour models have either of the two situations of (a) small left
mixing in $Y_\nu$ or (b) large left mixing in $Y_\nu$ (which can also 
be understood 
as mixing from the charged lepton sector). In an $SO(10)$ GUT, where there is
a unification of $Y_\nu$ and $Y_u$, both these situations can be realised
by choosing appropriate Higgs representations. We have christened these
two cases as the CKM-case and the PMNS case respectively. These two cases
have the following relations between the Yukawa matrices, in the basis
where charged lepton and down quark mass matrices are diagonal
\begin{eqnarray}
Y_\nu &=& Y_u \;\; \textrm{ (CKM case)} \\
Y_\nu &=& U_{PMNS} Y_u^{\mathrm{diag}} \;\; \textrm{ (PMNS case)} \;.
\end{eqnarray}
Both the above situations can be realised in $SO(10)$ without spoiling
the relation between neutrino Yukawa and the top Yukawa. The CKM case
can be realised by a simple superpotential involving only
ten-plets \cite{dimohall}
\begin{eqnarray}
W_{SO(10)} &=& (Y_u)_{ij} {\bf 16}_i {\bf 16}_j {\bf 10}_u + 
                (Y_d)_{ii} {\bf 16}_i {\bf 16}_i {\bf 10}_d 
               \nonumber \\
&&             + (Y_R)_{ij} {\bf 16}_i {\bf 16}_j {\bf 126}
\end{eqnarray}
where $i$ and $j$ are generation indeces.
The PMNS case is a bit more complicated as it can come from either a
renormalisable or non-renormalisable couplings. For example, Chang,
Masiero and Murayama \cite{changmasmur} have proposed the following
superpotential
\begin{eqnarray}
W_{SO(10)} &=& (Y_u)_{ij} {\bf 16}_i {\bf 16}_j {\bf 10}_u + 
               (Y_d)_{ii} {\bf 16}_i {\bf 16}_i 
               {{\bf 45}~{\bf 10} \over M_{\mathrm{Planck}}}
\nonumber\\
     &&          + (Y_R)_{ij} {\bf 16}_i {\bf 16}_j {\bf 126}
\end{eqnarray}
which leads to the PMNS like situation. Note that both the superpotentials
we have mentioned here are just scenarios but not complete fermion mass
models. For our purposes, these two scenarios serve as
the \textit{benchmark} points in the see--saw parameter space.

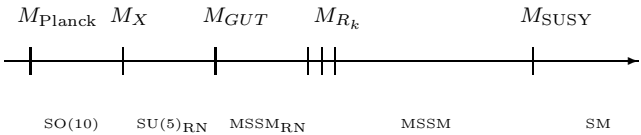
\begin{figure}[h]
{\footnotesize
\begin{center}
\begin{picture}(280,65)

\put(0,40){\vector(1,0){240}}

\put(5,55){$M_{\mathrm{Planck}}$}
\put(10,45){\line(0,-1){10}}
\put(40,55){$M_X$}
\put(45,45){\line(0,-1){10}}
\put(75,55){$M_{GUT}$}
\put(80,45){\line(0,-1){10}}

\put(117,55){$M_{R_k}$}
\put(115,45){\line(0,-1){10}}
\put(120,45){\line(0,-1){10}}
\put(125,45){\line(0,-1){10}}

\put(195,55){$M_{\mathrm{SUSY}}$}
\put(200,45){\line(0,-1){10}}

\put(15,15){\tiny SO(10)}
\put(50,15){\tiny  $\mathrm{SU(5)}_{\mathrm{RN}}$}
\put(85,15){\tiny $\mathrm{MSSM}_{\mathrm{RN}}$}
\put(150,15){\tiny MSSM}
\put(220,15){\tiny SM}
\end{picture}
\end{center}
}
\caption{\label{scales}Schematic
 picture of the energy scales involved in the model.}
\end{figure}

The LFV soft mass entries are generated in the RG evolution
from the universality scale down to the SUSY decoupling scale
$M_{SUSY}=1$ TeV. The breaking of SUSY $SO(10)$ down to
the SM  can be achieved in several different 
ways \cite{Mohapatrareview}. Within the two benchmark scenarios chosen
above, we envisage a breaking chain (Fig. \ref{scales}) of 
$SO(10) \to SU(5)_{\mathrm{RN}} \to \mathrm{MSSM_{RN}} \to \mathrm{MSSM}$.
Such a breaking can be achieved if the singlet under $SU(5)$ component 
of a ${\bf 16}$ or of a ${\bf 126}$ attains a VEV. 
 The scale of  $SU(5)_{\mathrm{RN}}$ is taken to be
the scale of the gauge coupling unification 
$M_{GUT}~ \sim~ 2~ \times 10^{16}$ GeV. The $SO(10)$ scale is considered
to be slightly higher about $M_{X}~ \sim~ 10^{17}$. This scale can be
considered to be the string unification scale, $M_{\mathrm{string}}$,
 which roughly 
turns out to be a factor $20-25$ from the gauge coupling unification scale 
after considering string loop effects \cite{dienesreview}. One interesting
aspect is that, while
we set the scale of the right handed neutrinos from low energy
neutrino data and $Y_\nu$ (as we will detail in section IV), it turns
out that the required right handed neutrino masses are close to the GUT
scale, which fits our scheme naturally.

Before proceeding in to next section where we detail the various lepton
flavour violating terms generated in these two schemes, we would like to
make some comments on the recent progress in $SO(10)$ model building. In the
recent years a new view regarding $SO(10)$ model building is being developed,
where construction of more realistic and complete models is being
pursued \cite{gorancharan}. In these `minimal' complete models, it is perhaps
for the first time possible to compute the entire $SO(10)$ spectrum, study
realistically precision observables such as running of fermion mass spectrum
including threshold effects, gauge couping running, proton decay,
etc \cite{charan2}. In the present work, we are more concerned
with the effect of $SO(10)$ see--saw couplings on the 
\textit{soft} supersymmetry
breaking sector of the theory. We do not resort to a complete model building
of $SO(10)$, but just consider schemes of $SO(10)$. 
This is sufficient for our
purposes, as we aim to compute the flavour violating entry in the slepton
and sneutrino mass matrices at the weak scale generated through the see--saw
mechanism within both these schemes. We  use 1-loop RGE equations for this
purpose and perform scatter plots in the SUSY breaking parameter space.\\

\section{LFV sources in SUSY--$SO(10)$}
In the present section, we  elaborate on the various contributions
to the lepton flavour violating entries in the $SO(10)$ SUSY--GUT
framework. Perhaps the
best way to understand them is in terms of the low-energy parameters. 
We  use the so--called Mass Insertion (MI) \cite{hallraby} 
notation to denote the
various flavour violating entries of the slepton mass matrix. These
flavour violating entries are zero at the high scale, where SUSY breaking
soft scalar masses are universal. At the weak scale, the universality
is broken by the RG evolution and the $6 \times 6$ slepton
squared-masses matrix $\mathcal{M}^2_{\tilde{\ell}}$
takes the form 
\begin{widetext}
\begin{equation}
\mathcal{M}_{\tilde{\ell}}^2 = 
\left(
\begin{array}{cc}
 m^2_{\tilde{\ell}}(1 + \delta_{LL}) 
+  Y_e Y_e^\dagger v_d^2 + \mathcal{O}(g^2) 
& v_d (A^\dagger_e - Y^\dagger_e \mu \tan\beta)  +
 \delta_{LR} \hat{m}^2_{\tilde{\ell}} \\
 v_d  (A_e -  Y_e \mu \tan\beta) + \delta_{RL} \hat{m}^2_{\tilde{\ell}}  &
m^2_{\tilde{\ell}}(1 + \delta_{RR})  + Y_e^\dagger Y_e v_d^2 + \mathcal{O}(g^2)
\end{array}
\right),
\label{smassdelta}
\end{equation}
\end{widetext}
where the flavour violation is coded in the $\delta$s given by
\begin{equation}
\label{smalldeltadef}
\delta_{ij} = \frac{\Delta_{ij}}{\hat{m}^2_{\tilde{\ell}}}
\end{equation}
with $\hat{m}^2_{\tilde{\ell}}$ being the geometric mean of the 
slepton squared masses \cite{gabbianimasiero88} and 
$\Delta_{i\ne j}$ flavour non-diagonal entries of the slepton mass
matrix generated at the weak scale by RG evolution. The 
mass insertions are divided in to the LL/LR/RL/RR types, according to
the chirality of the corresponding SM fermions. Detailed bounds on
each of these types of $\delta$s already exist in the literature
\cite{masina-savoy, paride}. 
Note that these bounds are obtained by considering one $\delta$ 
at time to 
be the source of the flavour violating effects, and putting all the
other $\delta$s to zero.
 
In our case, these $\Delta_{ij}$ are generated by RG evolution either
through the see--saw mechanism or through the GUT evolution. This means
that there exist several $\delta$s at the same time, so that
the interplay between them should be evaluated. However,
as an illustration, we  compare these resultant $\Delta$s generated
by RGEs with the existing MI bounds, conveying us the \textit{power} of each 
individual contribution at the weak scale in constraining the SUSY 
breaking parameter space. We  further elaborate the cases where 
double mass-insertions could be important compared to the single mass
insertions.

\begin{figure}[h]
\psfrag{m12}[c]{\huge{$M_{1/2}$ (GeV)}}
\psfrag{deltaLL}[c]{\huge{$(\delta_{12})_{LL}$}}
\psfrag{deltaRR}[c]{\huge{$(\delta_{12})_{RR}$}}
\psfrag{LL}[c]{\huge{$e\mu-LL$ insertion only}}
\psfrag{RR}[c]{\huge{$e\mu-RR$ insertion only}}
\psfrag{PRES}[r]{\Large{Present bound: BR$<1.2\cdot10^{-11}$}}
\psfrag{FUTU}[r]{\Large{Planned sensitivity: BR$\sim0.5\cdot10^{-13}$}}
\includegraphics[angle=-90, width=0.48\textwidth]{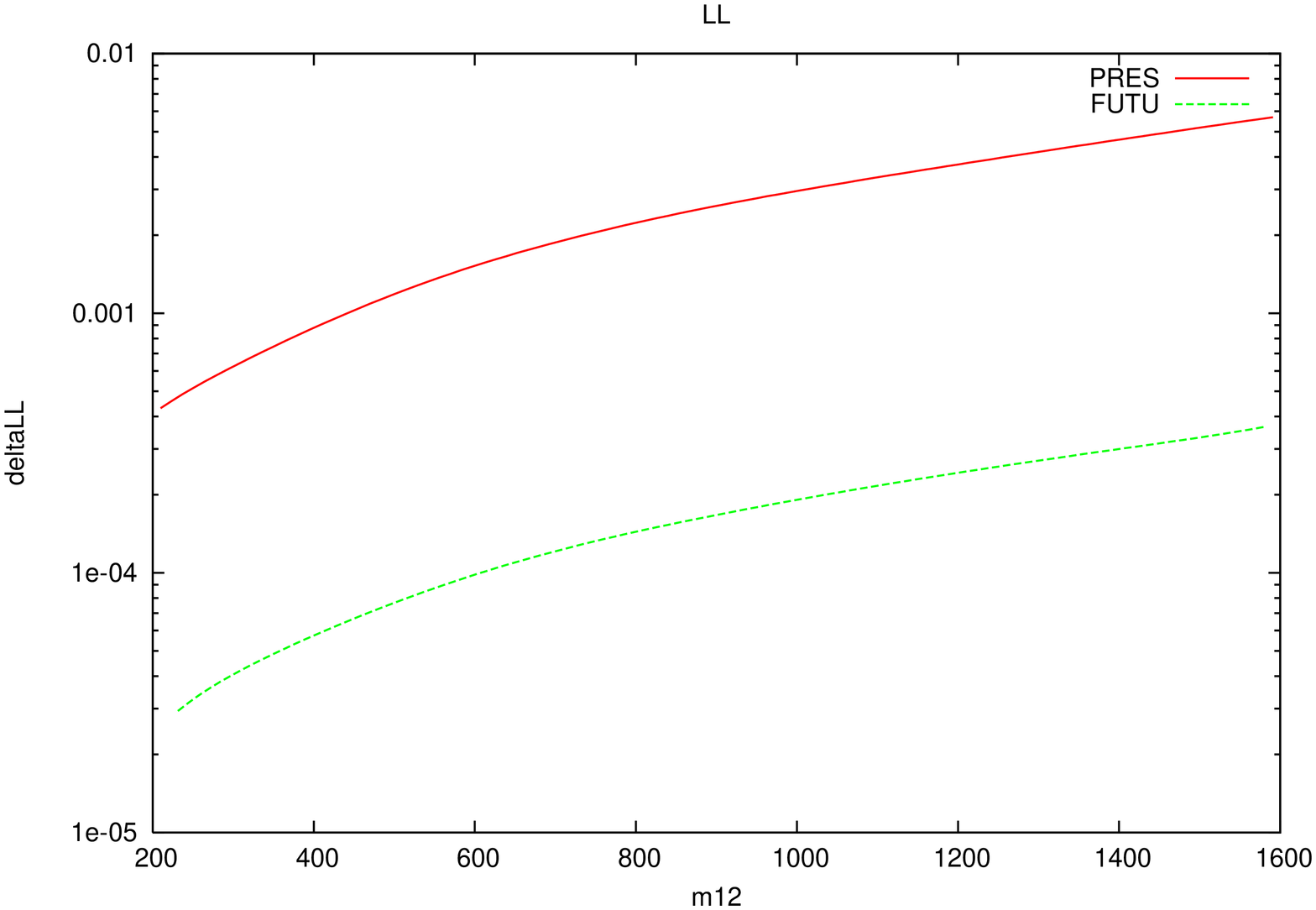}
\includegraphics[angle=-90, width=0.48\textwidth]{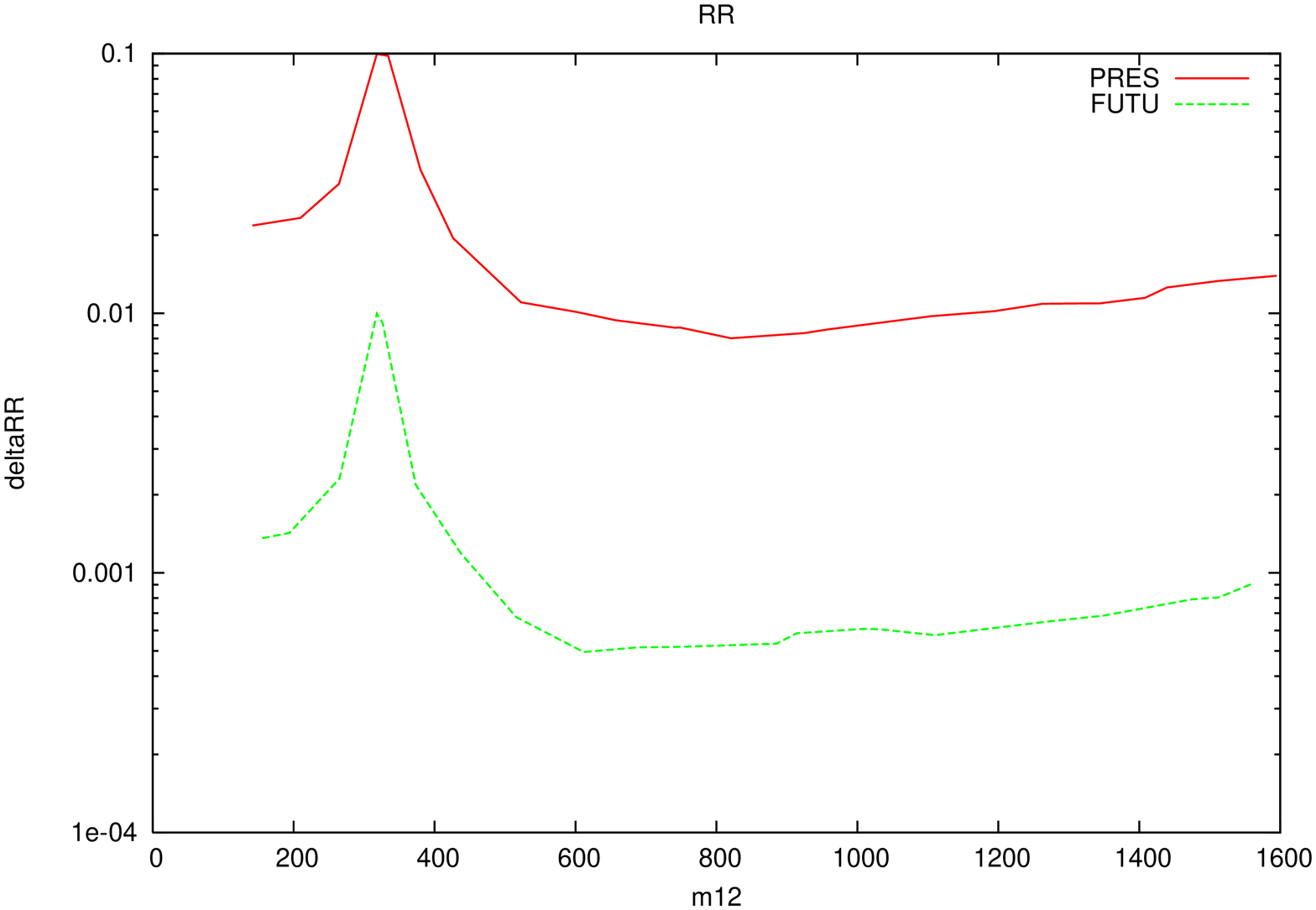}
\caption{\label{delta_graph}Points in the 
$(M_{1/2},(\delta_{12})_{LL})$   and 
$(M_{1/2},(\delta_{12})_{RR})$  planes that fulfill the
$\mu \rightarrow e\,\gamma$ branching ratio experimental limit. The plots
are for $\tan\beta=10$ and $m_0 = 500$ GeV.}
\end{figure}

For the discussion of this section we choose a point in the SUSY
parameter space:
$m_0 =500$ GeV; $M_{1/2}=500$ GeV; $A_0=0$; $\tan\beta=10$ (some
comments will be made also on an high $\tan\beta$ scenario, with
$t_\beta=40$). The SUSY spectrum for this point is given
in Table \ref{spectrum-tab}. As can be seen from 
Fig. \ref{delta_graph} there is no fine--tuned cancellation
in LFV amplitudes at this point, so that we can 
take it as a `safe' benchmark point. 

\begin{table}[b]
\caption{\label{spectrum-tab} Masses (in GeV) of the lightest SUSY particles
and of the higgs boson, corresponding to the SUSY--GUT point $m_0 = 500$ GeV,
 $M_{1/2} =500$ GeV, $A_0 = 0$.   }
\begin{ruledtabular}
\begin{tabular}{lcc}
Mass  & $\tan\beta=10$  & $\tan\beta=40$ \\
\hline
$m_{\tilde{\tau}_1}$  & 574 & 447     \\
$m_{\tilde{t}_1}$  & 845 & 838     \\
$m_{\tilde{g}}$  & 1225& 1225    \\
$m_{\tilde{\chi}^0_1}$  & 234 & 234     \\
$m_{\tilde{\chi}^+_1}$  & 431 & 432     \\
$m_{h}$  & 123 & 124     \\
\end{tabular}
\end{ruledtabular}
\end{table}

\begin{table*}[t!!!]
\caption{\label{table_delta}Bounds on the $\delta$s from the present and future
$BR(\ell_i \to \ell_j \, \gamma)$ experimental limits; by unconstrained we
mean that the $\delta$ is $\mathcal{O}(1)$.  
In the $\tau\mu$ sector the future sensitivity
is given both for SuperKEKB and for the proposed Super Flavour factory.
 The $\delta$s are calculated
at $t_\beta=10$, $m_0=500$ GeV and $M_{1/2}=500$ GeV; as can be
seen from Fig. \ref{delta_graph} no particular cancellation is
occurring at this point.}
\begin{ruledtabular}
\begin{tabular}{lcccccc}
& \multicolumn{2}{c}{LL} & \multicolumn{2}{c}{RR}
 & \multicolumn{2}{c}{LR}
\\ Process &Present & Future & Present & Future & Present & Future \\
\hline
$\mu\to e\,\gamma$ &
  $1.4\cdot10^{-3}$ & $8.5\cdot 10^{-5}$ & 
              $1.4 \cdot10^{-2}$ & $9\cdot 10^{-4}$ &
              $9.2 \cdot 10^{-6}$ & $5.9 \cdot 10^{-7}$\\
$\tau\to \mu\,\gamma$ &
  $2.4\cdot10^{-1}$ & 
$1.3\cdot 10^{-1}$/ $3\cdot10^{-2}$ & 
              uncon. & 
 uncon./ $2.9\cdot10^{-1}$       &
   $2.8 \cdot 10^{-2}$ &
 $1.5 \cdot 10^{-2}$ / $4.7 \cdot 10^{-3}$\\
$\tau\to e\,\gamma$ &
  $4.7\cdot10^{-1}$ & $1.4\cdot 10^{-1}$ & 
              uncon. & uncon. &
    $5.9 \cdot 10^{-2}$ & $1.5 \cdot 10^{-2}$
\end{tabular}
\end{ruledtabular}
\end{table*}

By turning on a $\delta$ at time, we get the present and
future bounds  on each  single
LFV $\delta$, as given in Table \ref{table_delta}.
From the table and from
Fig. \ref{delta_graph} it
is clear that the RR $\delta$s are less constrained 
than the LL ones by the LFV branching ratio bounds; 
this is due to two reasons: (i) the amplitudes involving
only $\delta_{RR}$ MI do not have chargino contributions
and (ii) in certain regions of the parameter space, there
could be cancellations between the bino and the 
higgsino--bino--higgsino contributions \cite{paride}.
It is also clear that the LR entries are much suppressed. 
This is mainly due to the
fact that, as can be seen by comparing  Eq. \eref{smassdelta}
and \eref{smalldeltadef}, in the normalization
procedure \eref{smalldeltadef} the $(\delta_{LR})_{ij}$
entry pays a factor ${m_i \over \hat{m}_{\ell}}$, 
where $m_i$ is the mass of the $i$-th lepton.

\subsection{LL insertions from the running}

To compute the $\Delta$'s from the RGEs, in this section
we  use the leading log approximation.  
Taking the soft masses to be flavour universal at the input scale, 
off diagonal entries in the LL sector are generated by right handed
neutrinos running in the loops; in our framework where $Y_{\nu_3} = Y_t$ 
we can estimate
\begin{equation}
(\Delta_{LL})_{i\neq j} = - \frac{3m^2_0 + A^2_0}{16\pi^2} Y^2_t V_{i3}V_{j3}
\ln \left(\frac{M^2_{X}}{M^2_{R_3}} \right)
\label{leadingloght}
\end{equation}
where $V$ can be either $V_{CKM}^T$ or $U_{PMNS}$, depending on the case.

\subsubsection{CKM case}

To use the leading log expression \eref{leadingloght}
we need to know the mass of the heaviest right handed neutrino.
By using the see--saw formula \eref{see-saw} we can estimate it
to be in the CKM case\cite{masivemp04}:
\begin{equation}
\label{mrapproxckm}
M_{R_3} \approx {m_t^2 \over 4~ m_{\nu_1}} \,;
\end{equation}
taking $m_{\nu_1} \approx 10^{-3}$ eV we get
$M_{R_3} \sim 10^{15}-10^{16}$ GeV.
The induced off-diagonal entries
relevant to $\ell_i \to \ell_j, \gamma$ are of the order of 
(putting $A_0$ to zero) 
\noindent 
\begin{eqnarray}
\label{wcmi1}
(\delta_{LL})_{\mu e}&=& -{3  \over 8 \pi^2}~
Y_t^2 V_{td} V_{ts} \ln{M_{X} \over M_{R_3}} 
\nonumber \\
\label{wcmi2}
(\delta_{LL})_{\tau\mu}&=& -{3 \over 8 \pi^2}~
Y_t^2 V_{tb} V_{ts} \ln{M_{X} \over M_{R_3}} 
 \\
\label{wcmi3}
(\delta_{LL})_{\tau e}&=& -{3 \over 8 \pi^2}~
Y_t^2 V_{tb} V_{td} \ln{M_{X} \over M_{R_3}} 
\, . \nonumber
\label{ckmleadinglog}
\end{eqnarray}
In these expressions, the CKM angles are small but one would expect 
the presence of the large top Yukawa coupling $Y_t$ to compensate such a 
suppression, giving rise to sizable $\delta$s.
We see from Table \ref{estimateCKM} that all the $\delta$s will be
outside the reach of planned experiments. Let us note that
the $\mu e$ sector entry is almost at the boundary of MEG
sensitivity: from Fig. \ref{delta_graph} it is clear that MEG
will test it for $M_{1/2} \lesssim 250$ GeV. 

\begin{table}[h]
\caption{\label{estimateCKM}CKM case: leading log estimates of
off-diagonal entries in the slepton mass matrices. The bounds
are calculated at $\tan\beta=10$, $m_0=500$ GeV and $M_{1/2}=500$ GeV.
For the $\tau\mu$ sector we give the sensitivity for both SuperKEKB and
a Super Flavour factory}
\begin{ruledtabular}
\begin{tabular}{lccc}
gen.  &  $|\delta_{LL}|$  & Present bound & Future
sensitivity\\
\hline
$\mu e$ & $3.4 \cdot 10^{-5}$ 
& $1.4 \cdot10^{-3}$ & $8.5 \cdot 10^{-5}$
\\
$\tau \mu$ & $ 6.2 \cdot 10^{-3} $ 
& $ 2.4 \cdot 10^{-1} $ & $1.3 \cdot 10^{-1} \:/\: 3.0 \cdot 10^{-2}$
\\
$\tau e $& $ 8.5 \cdot 10^{-4} $
& $ 4.7 \cdot 10^{-1} $ & $1.4 \cdot 10^{-1}$
\end{tabular}
\end{ruledtabular}
\end{table}

\subsubsection{PMNS case}
In the PMNS case the $R$ matrix is the identity; the see--saw formula
\eref{see-saw} can be straightforwardly inverted to get
\begin{equation}
\label{mrapproxmns}
\hat{M}_{R} = 
\mathrm{diag}\left\{ {m_u^2 \over m_{\nu_1}},~{m_c^2 \over m_{\nu_2}},
~{m_t^2 \over m_{\nu_3}} \right\}. 
\end{equation} 
Taking the neutrino spectrum to be hierarchical so that 
$m_{\nu_3} \approx \sqrt{\Delta m^2_{\rm atm}}$ we can estimate the
third right handed neutrino to have mass $M_{R_3} \sim 10^{14}$.
Plugging the value in the equation \eref{leadingloght}
\begin{eqnarray}
\label{mns1}
(\delta_{LL})_{\mu e}&=& -{3  \over 8 \pi^2}~
Y_t^2 U_{e3} U_{\mu3} \ln{M_X \over M_{R_3}} 
\nonumber \\
\label{mns2}
(\delta_{LL})_{\tau \mu}&=& -{3 \over 8 \pi^2}~
Y_t^2 U_{\mu3} U_{\tau3} \ln{M_X \over M_{R_3}} 
\\
\label{mns3}
(\delta_{LL})_{\tau e}&=& -{3 \over 8 \pi^2}~
Y_t^2 U_{e3} U_{\tau3} \ln{M_X \over M_{R_3}} 
\nonumber
\label{mnsleadinglog}
\end{eqnarray}
and taking $U_{e3}=0.07$ at about half of the current CHOOZ bound  we
get the estimates in Table \ref{estimateMNS}. We see from 
the table that the $\mu e$ sector is already ruled out
by the present bound and that the upcoming bound will be able
to test it up to
\begin{equation}
U_{e3} = 0.07 \cdot {8.5 \cdot 10^{-5} \over 1.4 \cdot 10^{-3}}
\sim 10^{-3} . 
\end{equation}
Moreover the $\tau \mu$ sector will be  probed by the SuperKEKB machine
 and thoroughly
tested by the proposed Super Flavour factory. 

\begin{table}[h]
\caption{\label{estimateMNS}  PMNS case: leading log estimates of
off-diagonal entries in the slepton mass matrices.
The values are for $m_0=500$ and $M_{1/2}=500$ GeV, $U_{e3}=0.07$ 
and $\tan\beta=10$. In $\tau\mu$ sector we give the sensitivities for 
both SuperKEKB and a Super Flavour factory.}
\begin{ruledtabular}
\begin{tabular}{lccc}
gen.  &  $|\delta_{LL}|$   & Present bound &
 Future sensitivity \\
\hline
$\mu e$ & $1.2\cdot 10^{-2}$ 
& $1.4 \cdot10^{-3}$ & $8.5 \cdot10^{-5}$
\\
$\tau \mu$ & $ 1.2 \cdot 10^{-1} $  
& $ 2.4 \cdot 10^{-1} $ & $1.3 \cdot 10^{-1} \:/\: 3.0 \cdot 10^{-2}$
\\
$\tau e$ & $ 1.2 \cdot 10^{-2} $ 
& $ 4.7 \cdot 10^{-1} $ & $1.4 \cdot 10^{-1}$
\end{tabular}
\end{ruledtabular}
\end{table}

\subsection{LR/RL insertions from the running}

The flavour violating terms in the LR sector are given
by the off--diagonal terms of the slepton' soft trilinear $(A_e)_{ij}$;
the RG generated entries in \eref{smassdelta} are
\begin{eqnarray}
(\Delta_{LR})_{i\neq j} 
&=& \langle H_d \rangle \big[ (A_e)_{ij}(M_X \to M_{GUT})\nonumber \\
&&+ 
   (A_e)_{ij}(M_{GUT}\to M_R) \big]\nonumber\\
&=& -\frac{3 m_i A_0}{32 \pi^2} \sum_k  Y_{\nu \: ik} 
Y_{\nu \: kj}^\dagger
\ln \left(\frac{M^2_{X}}{M^2_{R \: k}} \right)  \\
&& -\frac{9 m_j A_0}{32 \pi^2} \sum_k  Y_{u \: ik} 
Y_{u \: kj}^\dagger
\ln \left(\frac{M^2_{X}}{M^2_{GUT}} \right) \nonumber
\label{llLRall}
\end{eqnarray}
where $m_i$ is the mass of the $i$-th lepton
and the last line is coming from the fact that the left handed
sleptons and the $d^c$ squarks are hosted together
in the ${\bf 5}$ of $SU(5)$.
Taking into account only the third generation order one Yukawa
coupling we have
\begin{eqnarray}
(\Delta_{LR})_{ij} &=& -\frac{3 A_0}{32 \pi^2}
\Bigg[
 m_i Y_{\nu \: i3} Y^*_{\nu \: j3} 
\ln\left( \frac{M^2_X}{M_{R_3}^2}\right) \nonumber\\
&&+ 3m_j Y_{u \:  i3} Y^*_{u \: j3} 
\ln\left( \frac{M^2_X}{M_{GUT}^2}\right) 
\Bigg]
\end{eqnarray}
so that the see--saw driven contribution and the GUT
driven one give the dominant contribution to
different (transposed) entries. Let us note that
these entries are roughly equal, as 
the color factor 3 is almost compensated by the longer
running of the see--saw driven entries. As a
consequence, the remarks we are going to do about
the $ij$ entry will at the same time apply to
the transposed $ji$ one. 
Doing a comparison with \eref{leadingloght} we have
\begin{equation}
(\Delta_{LR})_{ij} = \frac{3 m_i A_0}{3 m^2_0 + A^2_0} (\Delta_{LL})_{ij}
\end{equation}
switching to the adimensional $\delta$s we get
\begin{equation}
|(\delta_{LR})_{ij}| = \frac{3|a_0|}{3+a^2_0} 
 \frac{m_i}{m_0} |(\delta_{LL})_{ij}| < \frac{m_i}{m_0} |(\delta_{LL})_{ij}|\:,
\label{smalldeltaLR}
\end{equation}
where $a_0 = A_0/m_0$.
A crucial point is that the RG generated  LR insertion 
\eref{smalldeltaLR} is not
the main contribution to the LR flavour violating insertion 
\cite{hisanomutanbeta}. Indeed
it is possible to build  an effective LR flavour violating insertion by
combining together the LL RG generated flavour violating entry 
\eref{leadingloght} with a 
LR flavour conserving $m_i \mu \tan\beta$ chirality flip
\begin{equation}
(\delta_{LR})^{\mathrm{eff}}_{ij}= {m_i \mu \tan\beta \over 
\hat{m}^2_{\tilde{\ell}} } (\delta_{LL})_{ij} \:.
\label{deltaLReff}
\end{equation}
Comparing Eq. \eref{smalldeltaLR} to \eref{deltaLReff}
\begin{equation}
\left|\frac{(\delta_{LR})_{ij}}{(\delta_{LR})_{ij}^\mathrm{eff}}\right|
< \frac{\hat{m}^2_{\tilde{\ell}}}{m_0 |\mu| \tan\beta} 
\approx (\tan\beta)^{-1}
\end{equation}
we see that
the effective LR insertion stemming from the RG generated LL one is 
enhanced by a factor $\tan\beta$ with respect to the RG
generated pure
LR insertion, so that the effective insertion
always dominates.

\subsection{RR insertions from the running}

The $SU(5)_{\mathrm{RN}}$ running from the soft breaking scale $M_X$
 to the GUT scale already breaks the
 universality of the soft spectrum by generating LFV entries at $M_{GUT}$. The 
renormalization group equation for $SU(5)_{\mathrm{RN}}$
 are calculated at the 1-loop level
in the  Appendix; the most interesting consequence of the $SU(5)_{\mathrm{RN}}$
running is that, as
both $Q$ and $e^c$ are hosted in the {\bf 10}, the CKM matrix mixing
the left handed quarks will give rise to off diagonal entries in
the running of the right handed slepton soft masses 
\cite{barbierihall, masip, hisanosu5, hisanonomura,%
masigut}
\begin{eqnarray}
(\Delta_{RR})_{i\neq j}& =& (m^2_{\tilde{10}})_{ij}(M_X 
\rightarrow M_{GUT}) \nonumber \\
&=& -3\cdot\frac{3m^2_0 + a^2_0}{16\pi^2} V_{ti}V_{tj}
\ln \left(\frac{M^2_X}{M^2_{GUT}} \right) , 
\end{eqnarray}
where we have used the fact that $Y_t \approx 1$ and $V_{tk}$,
$k=i,j$ is the $tk$ entry of the CKM matrix.
Let us note that this effect is due to the GUT structure, and is so
independent on any ansatz on the form of the neutrino Yukawa matrix:
in this sense the $\delta_{RR}$ and the BRs stemming from them form 
a guaranteed minimum on the LFV effects from a SUSY--GUT. 

\begin{table}[h]
\caption{\label{all_delta_table}Estimates of the $\delta$s from
leading log at $A_0=0$.}
\begin{ruledtabular}
\begin{tabular}{lccc}
gen. & $|\delta_{LL}|$ CKM & $|\delta_{LL}|$ PMNS & $|\delta_{RR}|$ \\
\hline
$\mu e$ & $ 3.4 \cdot 10^{-5}$ & $1.2 \cdot 10^{-2}$ &
$7.8 \cdot 10^{-5}$ 
\\ 
$\tau \mu$ &$6.2 \cdot 10^{-3}$ & $1.2 \cdot 10^{-1}$
&$1.4 \cdot 10^{-2}$ 
\\
$\tau e$  &$8.8 \cdot 10^{-4}$ & $1.2 \cdot 10^{-2}$
&$ 2 \cdot 10^{-3}$
\end{tabular}
\end{ruledtabular}
\end{table}

\begin{figure}[h]
\begin{center}
\psfrag{x}[c]{\huge{$M_{1/2}$ (GeV)}}
\psfrag{y}[c]{\huge{$BR(\mu \to e \, \gamma)\cdot 10^{11}$}}
\psfrag{deltas}[c]{\huge{Contributions to $BR(\mu \to e \, \gamma)$}}
\psfrag{FULL}[r]{\Large{Full CKM}}
\psfrag{LL}[r]{\Large{$(\delta_{LL})_{e\mu}$ only}}
\psfrag{RR}[r]{\Large{$(\delta_{RR})_{e\mu}$ only}}
\psfrag{DIA}[r]{\Large{$(\delta_{LL})_{e\tau}(\delta_{RR})_{\tau\mu}$ only}}
\psfrag{DIB}[r]{\Large{$(\delta_{LL})_{\mu\tau}(\delta_{RR})_{e\tau}$ only}}
\includegraphics[angle=-90, width=0.48\textwidth]{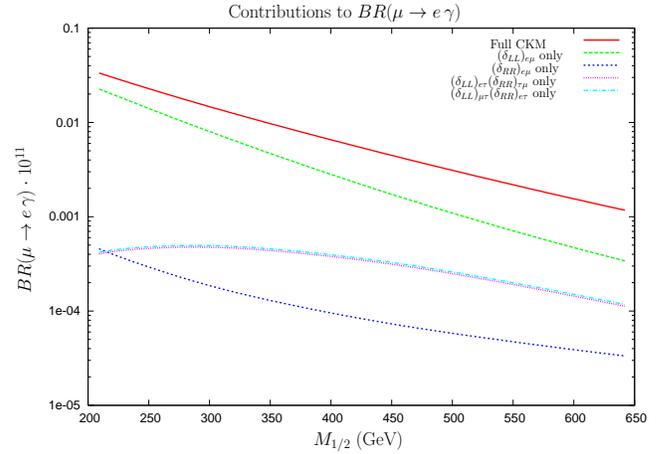}
\caption{\label{barbierihallfig}Comparison of the 
$\mu \to e \, \gamma$ BRs occurring from a full CKM case 
with those from just the LL entries, the  GUT generated RR entries
 and the double mass insertions.
The plots are done at $\tan\beta=40$, $m_0=500$ GeV and varying $M_{1/2}$
between 200 and 650 GeV. Details about the numerical procedure
will be given in the next section.}
\end{center}
\end{figure}

In  Table \ref{all_delta_table} we present a comparison of all
the $\delta$'s, other than the LR ones, for the two cases of minimal
and maximal mixing. We see  that in the PMNS case the 
main source of LFV violation are the LL
insertions, whereas in the CKM case the $SU(5)_{\mathrm{RN}}$
 running gives rise to a 
sizable right slepton off-diagonal mass
\begin{eqnarray}
(m^2_{\tilde{E}})_{i\neq j} &\approx& 2 (m^2_{\tilde{L}})_{i\neq j},
\end{eqnarray}
which could give a significant contribution to LFV amplitudes.
However,
as we mentioned at the beginning of this section, 
the $\delta_{RR}$ contribution to the
branching ratio is suppressed by the cancellations in the neutralino
sector. 
In Fig. \ref{barbierihallfig} we point out that the full BR in the
CKM case is of the same order of magnitude of the one calculated
by taking into account the LL entries only; on the
other hand, the ratio between
the $\delta_{LL}$ and the $\delta_{RR}$ generated BRs is more
than a factor 10. 

Following these results we can safely neglect the RR contributions and
estimate the branching ratios according to the formula \cite{hisanoformula}
(note that this already incorporates the effective LR insertion)
\begin{eqnarray}
BR(l_{i} \rightarrow l_{j} \gamma)&=&\frac{\alpha^{3}}{G_{F}^{2}}\cdot 
 \frac{(\delta_{LL})_{ij}^2}  {m_{SUSY}^{4}} \tan^{2}\beta
\label{hisano}
\\&\approx& 4.5 \cdot 10^{-6} \left( \frac{500 {\rm GeV}}{m_{SUSY}}\right)^4
(\delta_{LL})^2_{ij} \left( \frac{t_\beta}{10}  \right)^2
\nonumber
\end{eqnarray}
where $m_{SUSY}$ is linked to the high energy inputs parameters
$m_0$ and $M_{1/2}$ by the best fit relation \cite{petcovprofumo}
\begin{equation}
m^4_{SUSY}=0.5 \, \left( \frac{M^2_{1/2}}{m^2_0} \right)
\left(m^2_0 + 0.6 \, M^2_{1/2} \right)^2 \;.
\end{equation}
The estimated $BR(\ell_i \to \ell_j \, \gamma)$ are given in 
Table \ref{estimatedBRs}.  
We see that for $m_0$ and $M_{1/2}$ at 500 GeV the PMNS case is
already ruled out even at small $\tan\beta$ by the
$\mu\to e \, \gamma$ branching ratio, so that we expect the
upcoming MEG experiment to be able to test it even for soft masses
as big as 5 TeV. On the other
hand, we expect that the MEG  experiment will be able to test the
 small mixing angle case only for high values of $\tan\beta$.
As for the $\tau$ sector, we see that the only channel that offers
interesting rates is the $U_{e3}$ independent 
$\tau \to \mu \, \gamma$ process: the 
SuperKEKB bound of $10^{-8}$ will be able to
test the PMNS case even in the small $\tan\beta$ region, whereas
a Super Flavour factory, with a planned sensitivity of 
at least $\mathcal{O}(10^{-9})$,
is expected to address even the issue of small mixing angles,
provided that $\tan\beta$ is large.

\begin{table}
\caption{\label{estimatedBRs}Estimates of the branching ratios versus
present bounds and future sensitivities.
 $m_0$ and $M_{1/2}$ are taken to be 500 GeV.}
\begin{ruledtabular}
\begin{tabular}{lcccc}
gen. &  $t_\beta=40$ CKM & $t_\beta=10$ MNS & Exp. bound & Fut. sensit. \\
\hline
$\mu e$ & $ 6 \cdot 10^{-14}$ & $5 \cdot 10^{-10}$ &
$1.2\cdot 10^{-11}$ & $10^{-13} - 10^{-14}$ 
\\ 
$\tau \mu$ &$ 2\cdot 10^{-9}$ & $5 \cdot 10^{-8}$
&$6.8\cdot 10^{-8}$ & $10^{-8}$ 
\\
$\tau e$  &$ 4 \cdot 10^{-11}$ & $5 \cdot 10^{-10}$
&$3.1\cdot 10^{-7}$ & $10^{-8}$ 
\end{tabular}
\end{ruledtabular}
\end{table}

It is interesting to note from Fig. \ref{barbierihallfig} 
that the subleading contribution to the $\mu \to e \,\gamma$ 
process is not 
arising from a pure $(\delta_{RR})_{e\mu}$ insertion but from the double
FV mass insertion 
$(\delta_{LL})_{e\tau}(\delta_{RR})_{\tau\mu}+(e\leftrightarrow \mu)$:
this is an effective LR MI, which is enhanced by the flavour conserving
$m_{\tau} \mu \tan\beta$ contribution. 
This allows us
to give a rough estimate for the subleading contribution to the BR to
be
\begin{equation}
BR(\mu \to e \, \gamma)_{2\delta}
= \left({h_\tau \over h_\mu}\right)^2 
(\delta_{RR})^2_{\mu 3} BR(\tau \to \mu \, \gamma),
\label{BR2delta}
\end{equation}
where the suffix $2\delta$ represents the 2 flavour violating effective
MI.

\section{Numerical analysis of LFV processes}

\subsection{Parameter space of SUSY--GUTs}

As mentioned above, we  consider mSUGRA boundary conditions
for the soft masses. At the high scale, the parameters of the
model are the universal scalar mass $m_0$, universal trilinear
couplings $A_0$, unified gaugino masses $M_{1/2}$. In addition
to these there are the two Higgs potential parameters 
$\mu$ and $B$ and the undetermined ratio of the Higgs
VEVs, $\tan\beta$. The entire supersymmetric mass spectrum
is determined once these parameters are given. However, all these
parameters are not independent. Incorporating electroweak symmetry 
breaking gives rise to two conditions, reducing the number of 
independent parameters by two. In our case, we determine  
$\mu$ and $B$  by electroweak symmetry breaking conditions. 
The two conditions of the electroweak symmetry breaking are
\begin{eqnarray}
\label{mu0}
|\mu|^2 &=& \frac{m^2_{H_d} - m^2_{H_u} \tan^2\beta}{\tan^2\beta -1}
-\frac{1}{2}m^2_Z \nonumber \\
\label{bmu0}
\sin 2 \beta &=& {2 B \mu \over m_{H_u}^2 + m_{H_d}^2 + 2 \mu^2},
\end{eqnarray}
where $m_{H_u}^2$ and $m_{H_d}^2$ are the up and down type Higgs 
soft mass squared parameters determined at the weak scale, 
using the RG equations from the $M_{X}$ scale to the weak scale.
 At the weak scale, all
the supersymmetric soft parameters are thus known, enabling us to
compute the complete supersymmetric mass spectrum. 

We impose two main `theoretical' constraints on the SUSY 
parameter space: (a) Radiative ElectroWeak Symmetry Breaking 
(REWSB) \cite{ibanez} should take place.
 (b) No tachyonic particles and that the Lightest Supersymmetric
Particle (LSP) should be a neutralino \footnote{We do not impose
any additional constraints requiring unification of the Yukawa couplings 
in the present work.}. The experimental constraints are detailed in 
the next subsection. In contrast to the MSSM, both these constraints 
are significantly modified in the $SO(10)$ framework we discuss in the 
present paper. As it is well known, in the MSSM, radiative electroweak
symmetry breaking is driven by the top Yukawa coupling. In the $SO(10)$
model we are considering, two further effects are present: (i) the
range of the logarithmic running is a bit larger as $M_X$ is taken
to be $\sim 5 \times 10^{17}$ compared to typical MSSM studies, 
which consider the scale to be $\sim 2\times 10^{16}$; (ii) the
neutrino Yukawa couplings $Y_\nu$, one of which is 
necessarily as large as the
top Yukawa coupling, also contribute to driving the up--type Higgs 
soft mass squared
$m^2_{H_u}$ negative in the
running from the scales $M_X$ to $M_{R_3}$. These two contributions 
can significantly alter the parameter space which is viable under 
the electroweak breaking constraint. 

\begin{figure}[h]
\psfrag{y}[c]{\huge{$M_{1/2}$ (GeV)}}
\psfrag{x}[c]{\huge{$m_0$ (GeV)}}
\psfrag{mssm}[c]{\huge{MSSM parameter space}}
\psfrag{su5}[c]{\huge{SUSY--GUT parameter space}}
\psfrag{stop}[r]{\Large{$m(\tilde{t})=2.5$TeV}}
\psfrag{gluino}[r]{\Large{$m(\tilde{g})=2.5$TeV}}
\includegraphics[angle=-90, width=0.48\textwidth]{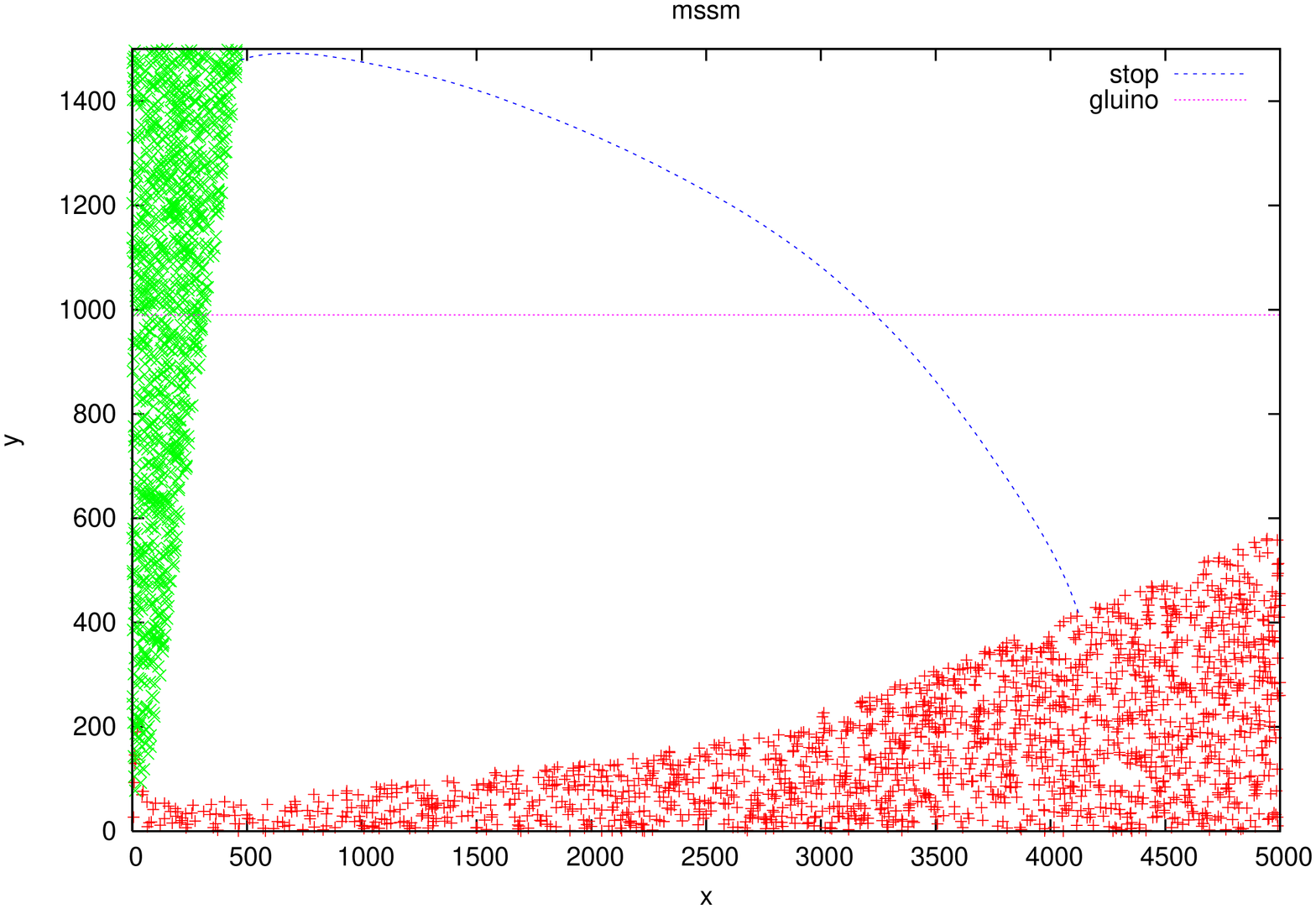}
\includegraphics[angle=-90, width=0.48\textwidth]{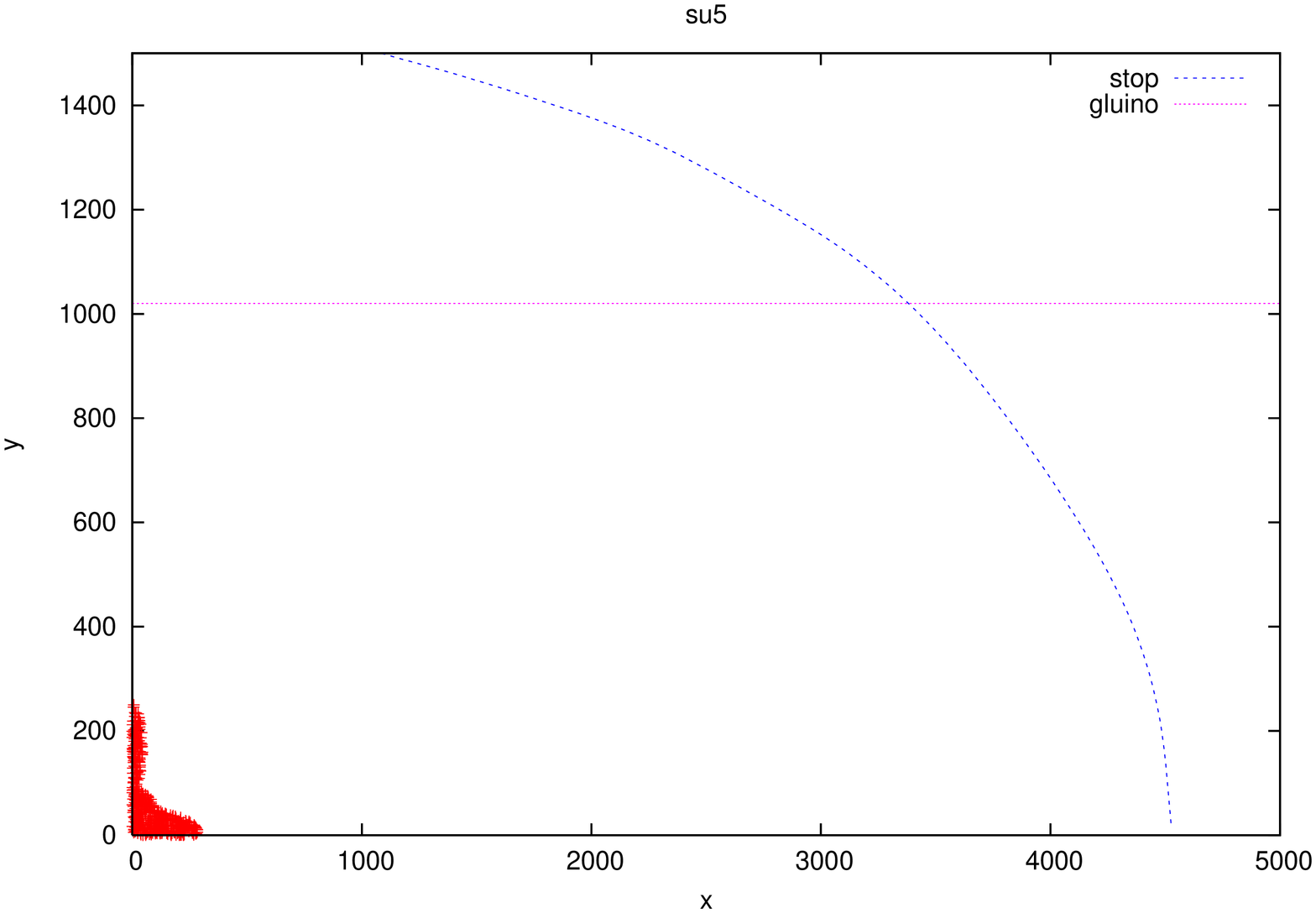}
\caption{\label{parspaces}Comparison of MSSM and SUSY--GUT parameter spaces.
 The colored areas are ruled out: green one correspond to points where the LSP
is not a neutralino; red ones to points where the vacuum is not viable (either
because of no REWSB or tachyonic particles)
The plots are for $\tan\beta=30$, $A_0 = 0$ and $m_t=173$ GeV.}
\end{figure}

A similar effect takes place for the region of the parameter space
in which the lightest slepton, which is typically the right-handed
stau $\tilde{\tau}_R$, is the LSP. In contrast to MSSM, in the
GUT framework, the stau also receives corrections from the 
`pre--GUT scale' running from the scale $M_{X}$ to the $M_{GUT}$.
In $SU(5)$, as we consider in the present scenario, the stau 
sits in the ten-plet ${\bf 10}$ which also hosts the strongly interacting
sector, leading to `strong' contributions through the gaugino loops. 
A leading log estimate of these contributions for $m_0 \approx 0$ is given by 
\begin{equation}
m^2_{\tilde{\tau}_R}(M_{GUT}) = \frac{96}{80 \pi^2} 
M^2_{1/2}\ln
\left( \frac{M_X}{M_{GUT}} \right) \approx 0.4 M^2_{1/2} \;.
\label{stimamia}
\end{equation}
These positive contributions can thus off--set the negative contributions
from the Yukawa running. Both these effects are best demonstrated in the
Fig. \ref{parspaces}, where the difference in the allowed parameter
space of the MSSM and of the $SO(10)$ framework is evident.  

Finally, as mentioned in the introduction, we would like to do a 
complementarity
study between the region of the parameter space probed at the LHC 
{\it vis-a-vis}
the LFV experiments. For this, we first need to determine the
region of the parameter space probed by the LHC considering various 
detection channels, putting the appropriate background cuts, detector 
response functions etc. We do not intend to do such detailed analysis
in this work. For mSUGRA, it is already
present in the literature \cite{tata}. The typical estimate for the mass of the
gluino and squarks to be detected at the LHC is about 2-3 TeV. We define
the parameter space region that allows a  squark mass to 
be below 2.5 TeV to be the region probed by the LHC. In Fig. 
\ref{parspaces} the contours for the masses are shown in
the $(m_0$,$M_{1/2})$ plane. We call this region the LHC accessible
region. However, we also further consider other regions of the
parameter space which, though not accessible at the LHC, can be relevant
for the reach of flavour physics experiments. With this in mind, we scan 
the total parameter space in the following ranges 

\begin{eqnarray*}
m_0 &\in& (0,5000)\mathrm{\;\:GeV}\\
M_{1/2} &\in& (0,1500)\mathrm{\;\:GeV} \\
A_0 &\in& (-3\cdot m_0,+3 \cdot m_0) \\
\tan \beta&=& 10,40 \\
\mathrm{sign} \mu &\in& \{+,- \}
\end{eqnarray*}

\subsection{Integration procedure}

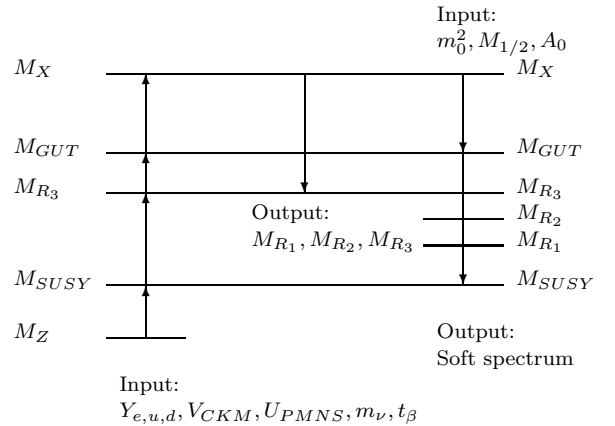
\begin{figure}[h]
{\footnotesize
\begin{center}
\begin{picture}(210,160)

\put(35,30){\line(1,0){30}}
\put(35,50){\line(1,0){150}}
\put(35,85){\line(1,0){150}}
\put(35,100){\line(1,0){150}}
\put(35,130){\line(1,0){150}}

\put(0,30){$M_Z$}
\put(0,50){$M_{SUSY}$} 
\put(190,50){$M_{SUSY}$}
\put(0,100){$M_{GUT}$}
\put(190,100){$M_{GUT}$}
\put(0,130){$M_X$}
\put(190,130){$M_{X}$}

\put(50,30){\vector(0,1){20}}
\put(50,50){\vector(0,1){35}}
\put(50,85){\vector(0,1){15}}
\put(50,100){\vector(0,1){30}}

\put(110,130){\vector(0,-1){45}}

\put(170,130){\vector(0,-1){30}}
\put(170,100){\vector(0,-1){50}}

\put(40,10){Input:}
\put(40,0){$Y_{e,u,d}, V_{CKM}, U_{PMNS}, m_{\nu}, t_\beta$}

\put(160,150){Input:}
\put(160,140){$m^2_0, M_{1/2}, A_0$}

\put(90,75){Output:}
\put(90,65){$M_{R_1}, M_{R_2}, M_{R_3}$}

\put(160,30){Output:}
\put(160,20){Soft spectrum}

\put(155,75){\line(1,0){30}}
\put(155,65){\line(1,0){30}}

\put(0,85){$M_{R_3}$}

\put(190,85){$M_{R_3}$}
\put(190,75){$M_{R_2}$}
\put(190,65){$M_{R_1}$}

\end{picture}
\end{center}
}
\caption{\label{routine} Pictorial explanation of the running routine.
See the text for the details.}
\end{figure}

In the present section, we detail the integration procedure we have
incorporated in our work. A schematic diagram is presented in 
Fig. \ref{routine}. As inputs at the weak scale, we consider
the Yukawa couplings of the up-type quarks, down-type quarks, charged
leptons, the CKM mixing matrix and tan$\beta$. We employ a hierarchical
scheme for the neutrino masses. The lightest neutrino is taken to be
around $10^{-3}$ eV. The other two neutrino masses are determined by
the square-roots of the solar and atmospheric mass squared differences
respectively. The leptonic mixing matrix $U_{PMNS}$ has two large 
mixing angles, and the unknown third mixing angle $U_{e3}$ is left
as a free parameter. Unless otherwise stated, we take
$U_{e3}=0.07$, 
half of the current upper limit from the CHOOZ experiment. 
We use 1-loop RGEs to run all the Yukawa couplings up to the high
scale. For the neutrino masses and mixing we use the RGE given
in the literature \cite{chankowski-pokorski, lindnerantusch}. 

As a first step, we run the neutrino mass matrix and
the Yukawa and gauge couplings up to the 
right handed neutrino masses, using an estimated $M_{R_3}$,
given by \eref{mrapproxckm} in the CKM case and by
\eref{mrapproxmns} in the PMNS one. At that scale we
assign the neutrino Yukawa matrix: in the CKM case evaluate it 
to be $Y_\nu = Y_u$; in the PMNS case we first extract 
$U_{PMNS}(M_{R_3})$  and then define 
$Y_\nu = U_{PMNS}(M_{R_3})Y^{\mathrm{diag}}_u$. 
We then run up to the $M_X$ scale and we redefine $Y_\nu(M_X)$ to be
equal to $Y_u(M_X)$ in the CKM case, or to $U_{PMNS}(M_{X})Y^{\mathrm{diag}}_u$ in the PMNS case. Once that the neutrino Yukawa 
matrix is known at $M_X$ we are able to use the see-saw formula 
(\ref{see-saw}) in order to calculate the right handed neutrinos 
mass matrix and, thus, the energies at which each heavy neutrino 
should decouple; we have only to use again the RGEs down to
the estimated $M_{R_3}$ and do the 
calculation \footnote{This last step is necessary only in the
CKM case, as the relations \eref{mrapproxmns} are exact at the scale
$M_R$.}.
We thus use the iterative method to check if our results are right. 

With this information, high energy inputs and the intermediate energy 
scales, we are now ready to compute the running of the soft spectrum
from the high scale to the weak scale. We do this using 1-loop
RGEs \cite{falck}. 
At the weak scale, we compute the full 6$\times$ 6 mass matrices
of all the scalars and the neutralino and chargino mass matrices.  
In the Higgs sector we employ the full 1--loop effective 
potential \cite{zwirner} to determine the parameters and compute
the spectrum. Finally we impose various direct experimental constraints
as well as the theoretical constraints on the SUSY parameter space:
\begin{itemize}
\item LEP mass limit on the lightest Higgs;
\item direct search limits on charginos and sfermions; 
\item neutral LSP;
\item viable vacuum: REWSB at $M_Z$ and no tachyonic particles.
\end{itemize}

For every point which passes through all these constraints, 
we compute leptonic flavour violating decay rates by using
the exact mixing matrices \cite{hisanonomura} as well as masses for the 
sleptons, neutralinos and charginos.

\section{Results}

From our leading log estimates we expect that the most promising
sectors for finding SUSY--GUT induced LFV
are the $\mu e$ and the $\tau \mu$ ones. 
Given that the planned sensitivities (Table \ref{lfvtable})
to all the LFV processes
will be of the same order ($\sim \mathcal{O}(10^{-13}-10^{-14})$
in the $\mu e$ sector and $\sim \mathcal{O}(10^{-8})$ in the
$\tau\mu$ one), we  concentrate on the two body
decays, $\mu \to e \, \gamma$, to be probed by the MEG experiment
at PSI, and $\tau \to \mu \, \gamma$, that is under study
at Beauty factories. Indeed, the three body decays 
 are weaker probes of SUSY--GUTs, as the leading
penguin contribution leads to a BR that is suppressed by
a factor $\sim \alpha$ with respect to the two body
decay. The $\mu \to e$ conversion in Nuclei process suffers from
a similar suppression, but due to the well defined
experimental signal the PRISM/PRIME aims to a huge improvement 
in the sensitivity to offset this factor.

In this section we  display the results from the
numerical routine for the processes of interest.
All the plots are done for positive
$\mu$ as there is no sensible difference with the negative $\mu$ case
as far as lepton flavour violating processes 
are concerned \footnote{Let us note that 
the $\mu<0$ scenario is strongly disfavored by bounds on the
FCNC $b \to s\, \gamma$ amplitude and by the SUSY corrections
to  $(g_\mu -2)$}.

\subsection{The MEG experiment at PSI}

\begin{figure}[h]
\psfrag{y}[c]{\huge{$BR(\mu \to e\,\gamma)\cdot10^{11}$ }}
\psfrag{x}[c]{\huge{$M_{1/2}$ (GeV)}}
\psfrag{title10}[c]{\huge{$\mu\to e\,\gamma$ at $\tan\beta=10$}}
\psfrag{title40}[c]{\huge{$\mu\to e\,\gamma$ at $\tan\beta=40$}}
\psfrag{CKM}[c]{\Large{CKM-case}}
\psfrag{MNS}[c]{\Large{PMNS-case}}
\includegraphics[angle=-90, width=0.48\textwidth]{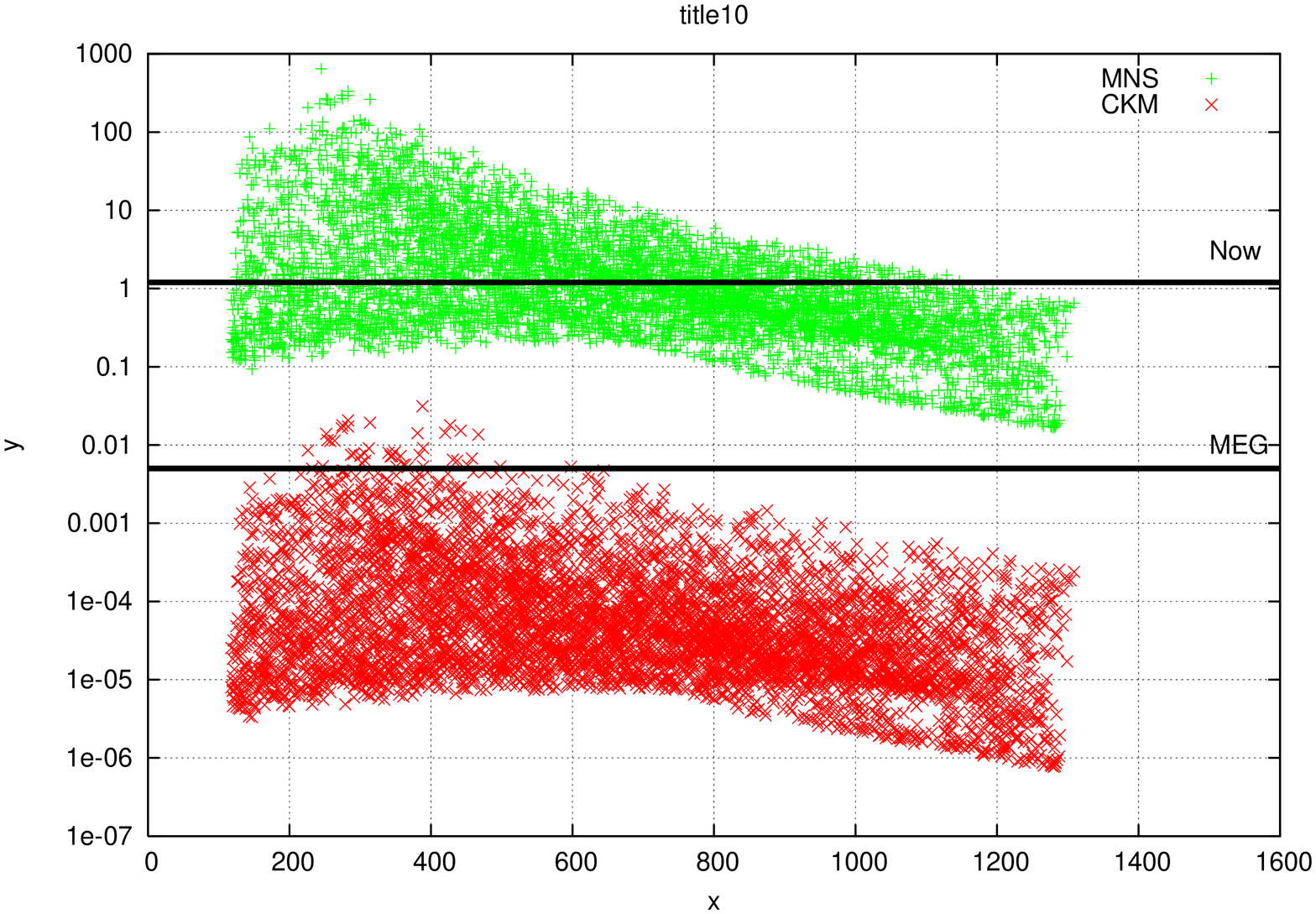}
\includegraphics[angle=-90, width=0.48\textwidth]{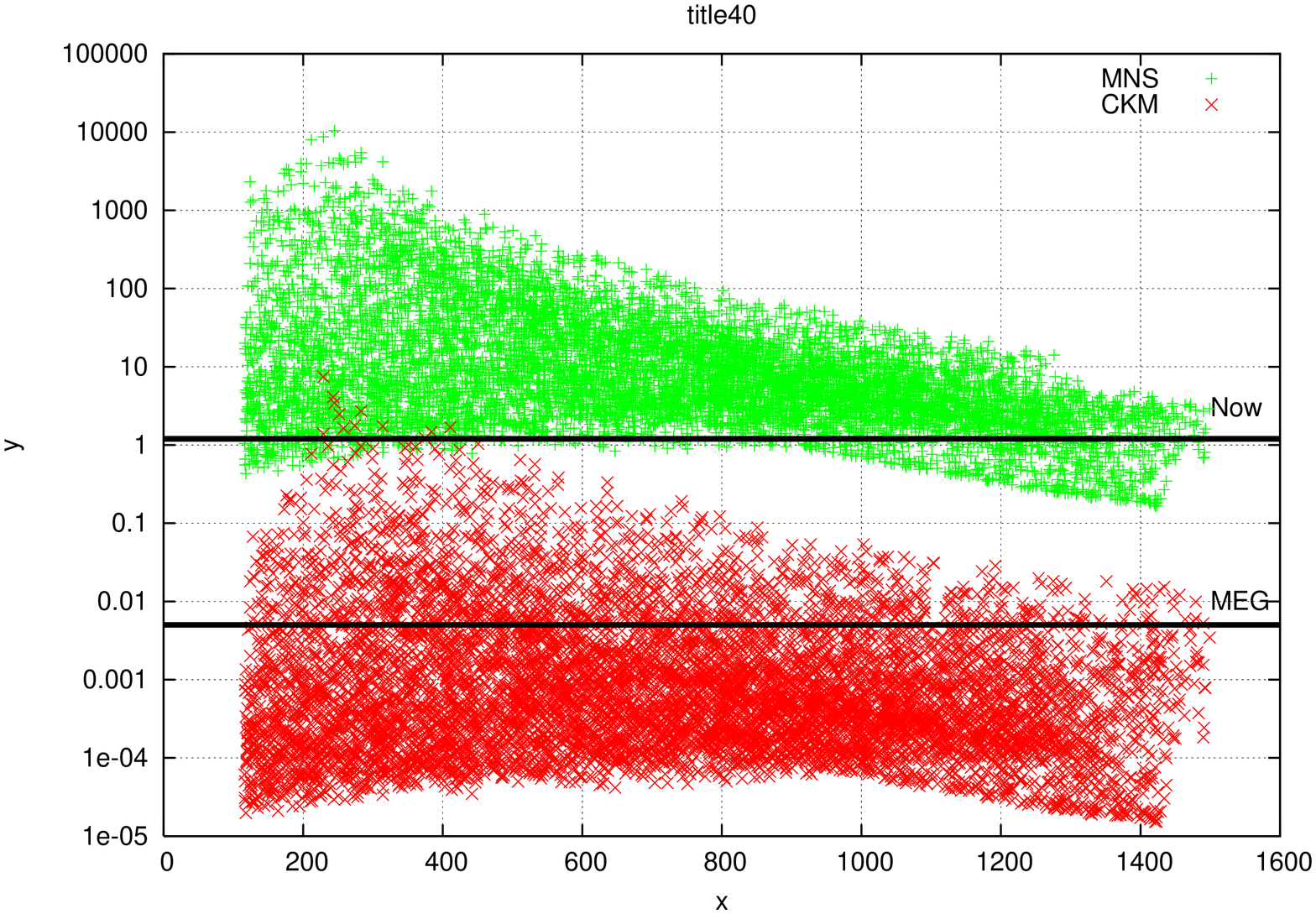}
\caption{\label{megplot}Scaled BR($\mu\to e\,\gamma$) vs. $M_{1/2}$.
The plots are obtained by scanning the LHC accessible
SUSY--GUT parameter space at fixed values of $\tan\beta$.
The horizontal
lines are the present (MEGA) and the future (MEG) experimental sensitivities.
Note that MEG will test the PMNS case and, for high $\tan\beta$,
 constrain the CKM one.}
\end{figure}

\begin{figure}[h]
\begin{center}
\psfrag{x}[c]{\huge{$m_0$ (GeV)}}
\psfrag{y}[c]{\huge{$M_{1/2}$ (GeV)}}
\psfrag{title10}[c]{\huge{$\mu\to e\,\gamma$ at $\tan\beta=10$}}
\psfrag{ckm40}[c]{\huge{CKM--case, $\tan\beta=40$}}
\psfrag{10E-13}[c]{\Large{$BR=10^{-13}$}}
\psfrag{10E-14}[c]{\Large{$BR=10^{-14}$}}
\includegraphics[angle=-90, width=0.48\textwidth]{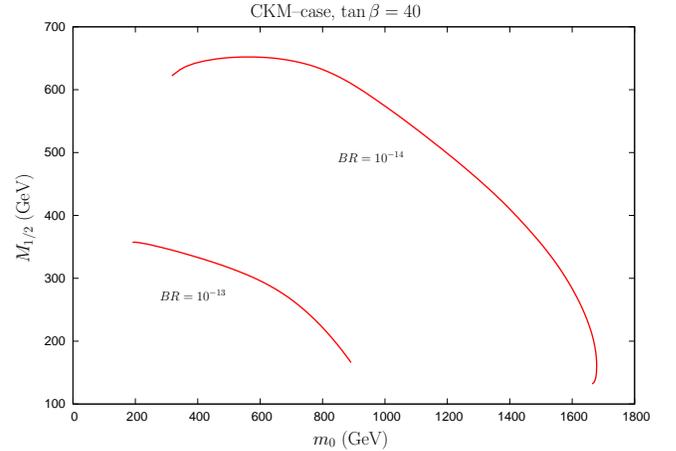}
\end{center}
\caption{\label{ckmcontour}Contour plots at fixed $BR(\mu \to e\,\gamma)$
in the $(m_0, M_{1/2})$ plane, at $A_0=0$ in a CKM high $\tan\beta$
case. Note that while the plane is presently unconstrained, the MEG
experiment sensitivity of $\mathcal{O}(10^{-13}-10^{-14})$
 will be able to probe it in the 
$(m_0, m_{\tilde{g}}) \lesssim 1$ TeV
region.}
\end{figure}

Given the planned astonishing sensitivity of the upcoming
MEG \cite{meg} experiment at PSI,
 we expect that the muon decay
$\mu \to e\, \gamma$ will be
a very interesting probe of LFV in a SUSY--GUT scenario.
This statement is quantified in 
Figs. \ref{megplot} and \ref{ckmcontour}: 
the PMNS case high $\tan\beta$ scenario is
already ruled out by the current MEGA \cite{mega}
 bound on the BR($\mu \to e \, \gamma$);
 the low $\tan\beta$ regime is already severely
constrained for not too high $M_{1/2}$ and will be completely
probed by the upcoming MEG experiment.
The CKM case, instead, is below the present bounds in all the
parameter space,  but a sensible portion of the high  
$\tan\beta$ regime will be within the reach of MEG sensitivity
(Fig. \ref{ckmcontour}).

\begin{table}[h!!!]
\caption{\label{megconc}Reach in $(m_0, m_{\tilde{g}})$ of the past
 and upcoming experiments from
their $\mu \to e \, \gamma$ sensitivity. LHC means that all the LHC testable 
parameter space will be probed; all means that soft masses
as high as $(m_0, m_{\tilde{g}}) \lesssim 5$ TeV will be probed.}
\begin{ruledtabular}
\begin{tabular}{lcccc}
&\multicolumn{2}{c}{PMNS}&\multicolumn{2}{c}{CKM}\\
Exp.
&$t_\beta=40$ & $t_\beta=10$ &$t_\beta=40$ & $t_\beta=10$
\\ \hline 
MEGA &
 LHC  & 2 TeV  & no & no
\\
MEG &
 all & all & 1.3 TeV  & no
\end{tabular}
\end{ruledtabular}
\end{table}

\begin{figure}[h]
\psfrag{y}[c]{\huge{$BR(\mu \to e\,\gamma)\cdot10^{11}$ }}
\psfrag{x}[c]{\huge{$M_{1/2}$ (GeV)}}
\psfrag{title10}[c]{\huge{$\mu\to e\,\gamma$ at $\tan\beta=10$}}
\psfrag{title40}[c]{\huge{$\mu\to e\,\gamma$ at $\tan\beta=40$}}
\psfrag{MNS}[c]{\Large{PMNS-case}}
\includegraphics[angle=-90, width=0.48\textwidth]{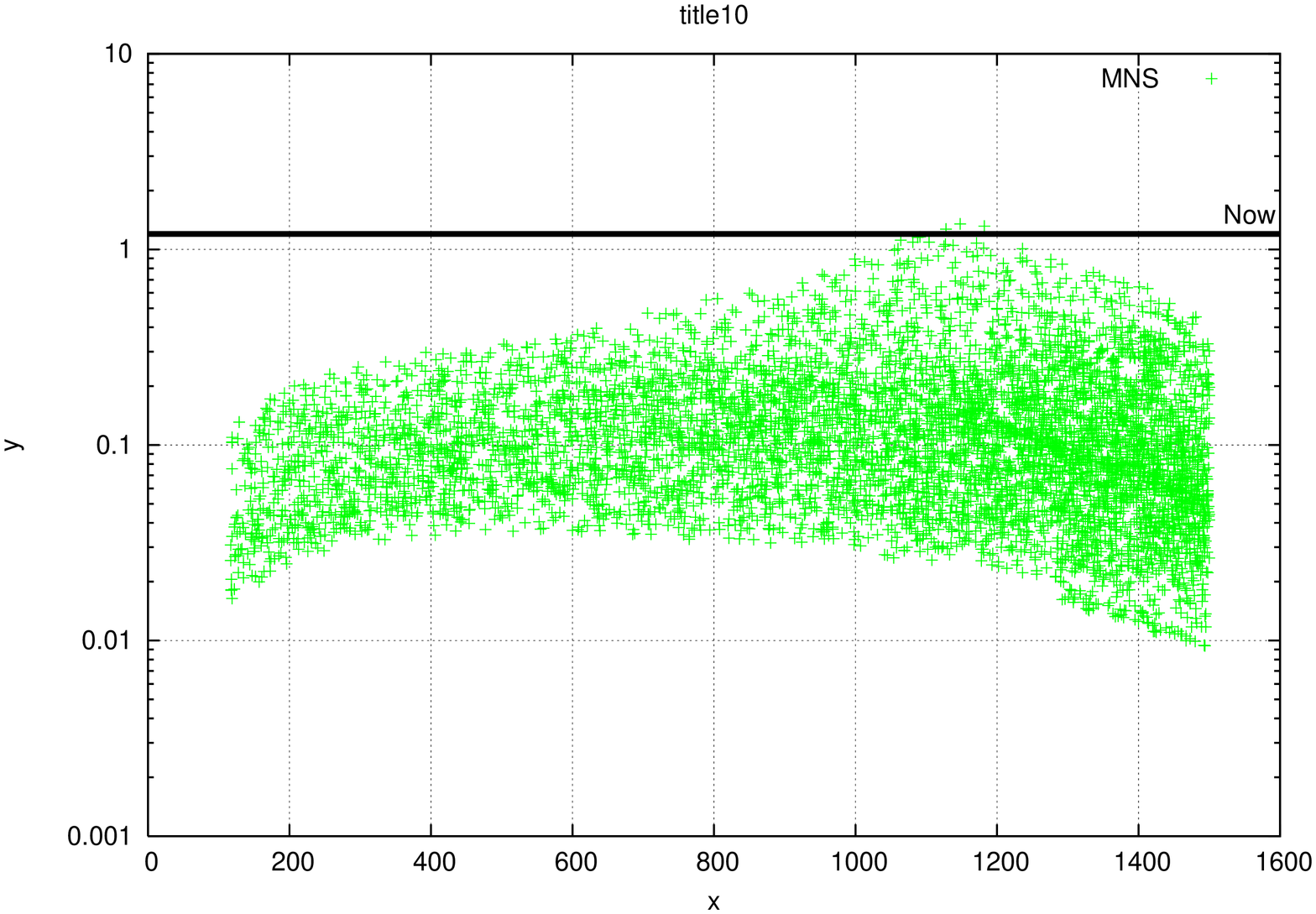}
\includegraphics[angle=-90, width=0.48\textwidth]{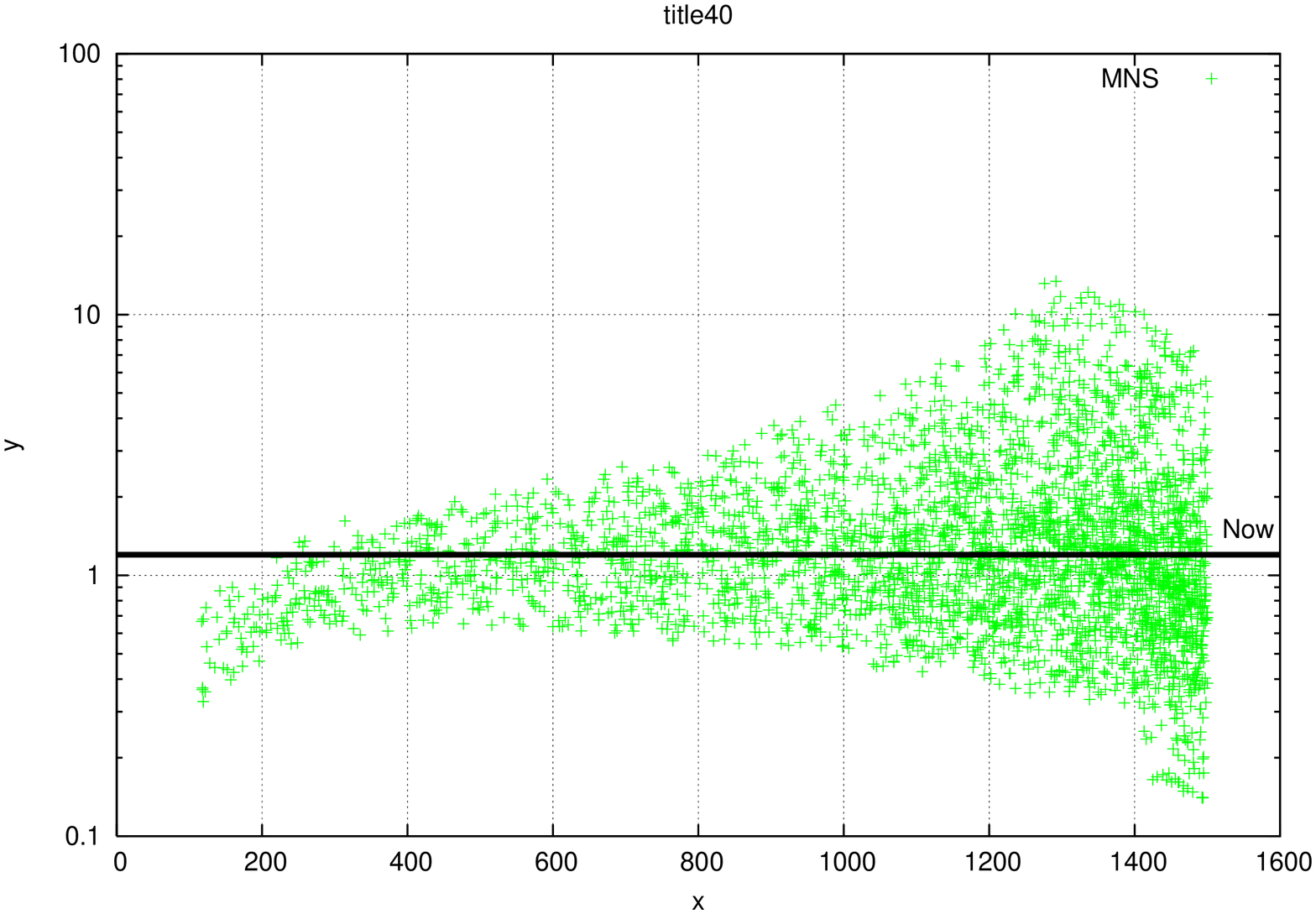}
\caption{\label{nodir}Scaled BR($\mu\to e\,\gamma$) vs. $M_{1/2}$ outside
LHC experiments' reach for low and high $\tan\beta$.
 The horizontal
line is the present MEGA bound. The upcoming MEG sensitivity
will test all the points.}
\end{figure}

This allows us to draw the conclusion that
(Table \ref{megconc}), for not too big values
of the soft breaking parameters 
(i.e.: $(m_0, m_{\tilde{g}}) \lesssim 1$ TeV), 
the MEG experiment
will be able to find evidence of SUSY induced lepton flavour violation,
unless we are in a low $\tan\beta$, small mixing SUSY--GUT:
as a consequence, if the LHC finds supersymmetry to be at the TeV scale
but $\mu \to e\,\gamma$ escapes MEG detection, this will be the
only viable SUSY $SO(10)$ see--saw scenario. 
Moreover, as depicted in Fig. \ref{nodir}, in the PMNS case
the sensitivity of MEG will outreach that of the LHC, 
being able to probe soft masses as high as
$(m_0 = 5, M_{1/2}=1.6)$ TeV - so that if MEG gets positive evidence
but the LHC fails in its aim to detect superpartners
the viable SUSY--GUTs will be restricted 
to the high soft mass regime with large mixing
angle in the neutrino Yukawa sector.

\subsection{B factories, SuperKEKB and Super Flavour factory}

The $\tau\mu$ sector poses  promising prospects of discovery of
SUSY--GUT induced lepton flavour violation in
the case that the planned Super Flavour factory
\cite{superflavour} will be realized:
let us note that this machine is planned to reach a
sensitivity of at least 
$BR(\tau \to \mu\,\gamma)\sim\mathcal{O}(10^{-9})$,
with an improvement of the present bound by nearly two
orders of magnitude. The main theoretical interest for such
process arise from the fact that the dominant LFV insertion
$(\delta_{LL})_{\tau\mu}$ does not depend on the unknown PMNS
angle $U_{e3}$.

\begin{figure}[t]
\psfrag{y}[c]{\huge{$BR(\tau \to \mu\,\gamma)\cdot10^{7}$ }}
\psfrag{x}[c]{\huge{$M_{1/2}$ (GeV)}}
\psfrag{title10}[c]{\huge{$\tau\to \mu\,\gamma$ at $\tan\beta=10$}}
\psfrag{title40}[c]{\huge{$\tau\to \mu\,\gamma$ at $\tan\beta=40$}}
\psfrag{CKM}[c]{\Large{CKM-case}}
\psfrag{MNS}[c]{\Large{PMNS-case}}
\includegraphics[angle=-90, width=0.48\textwidth]{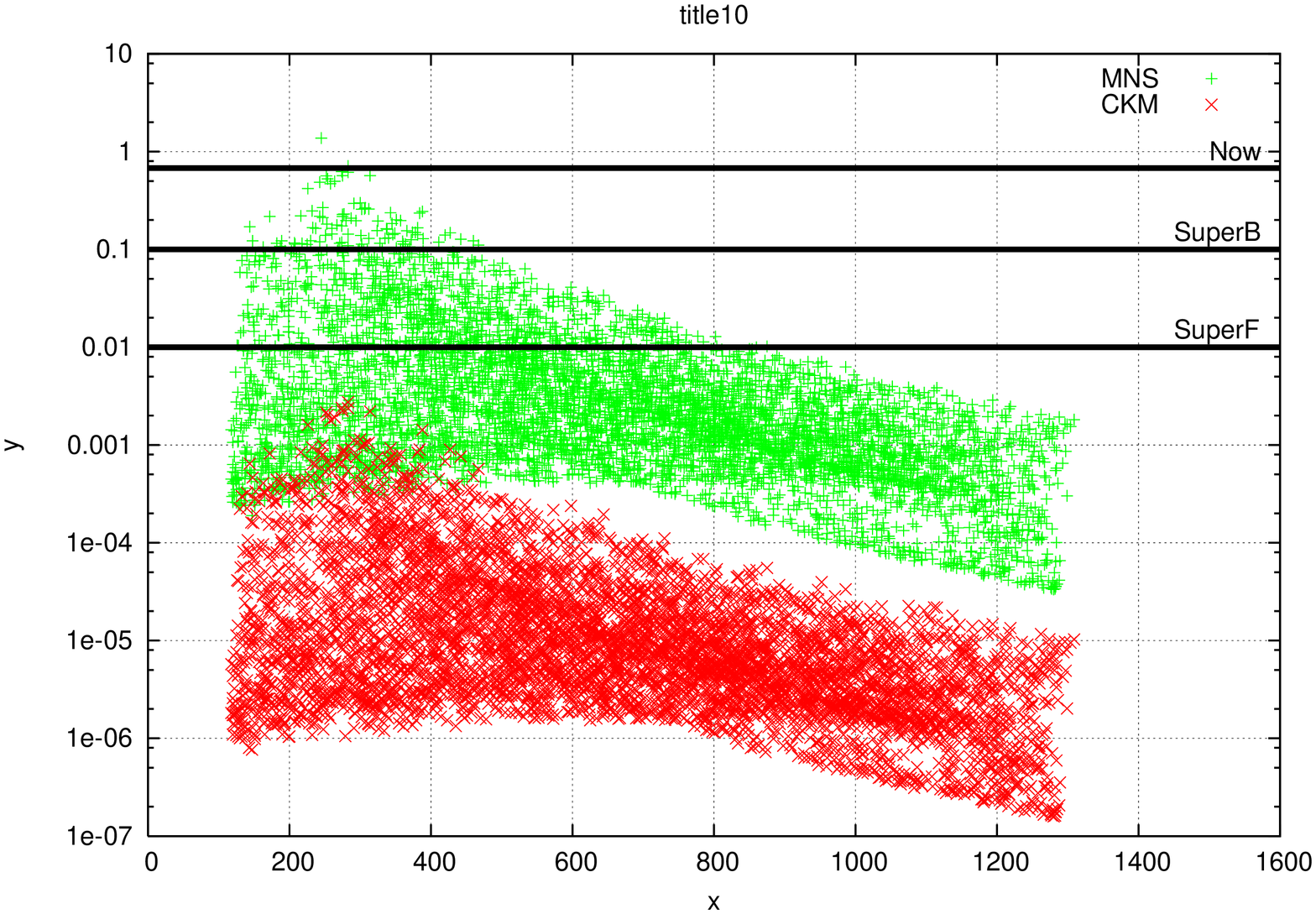}
\includegraphics[angle=-90, width=0.48\textwidth]{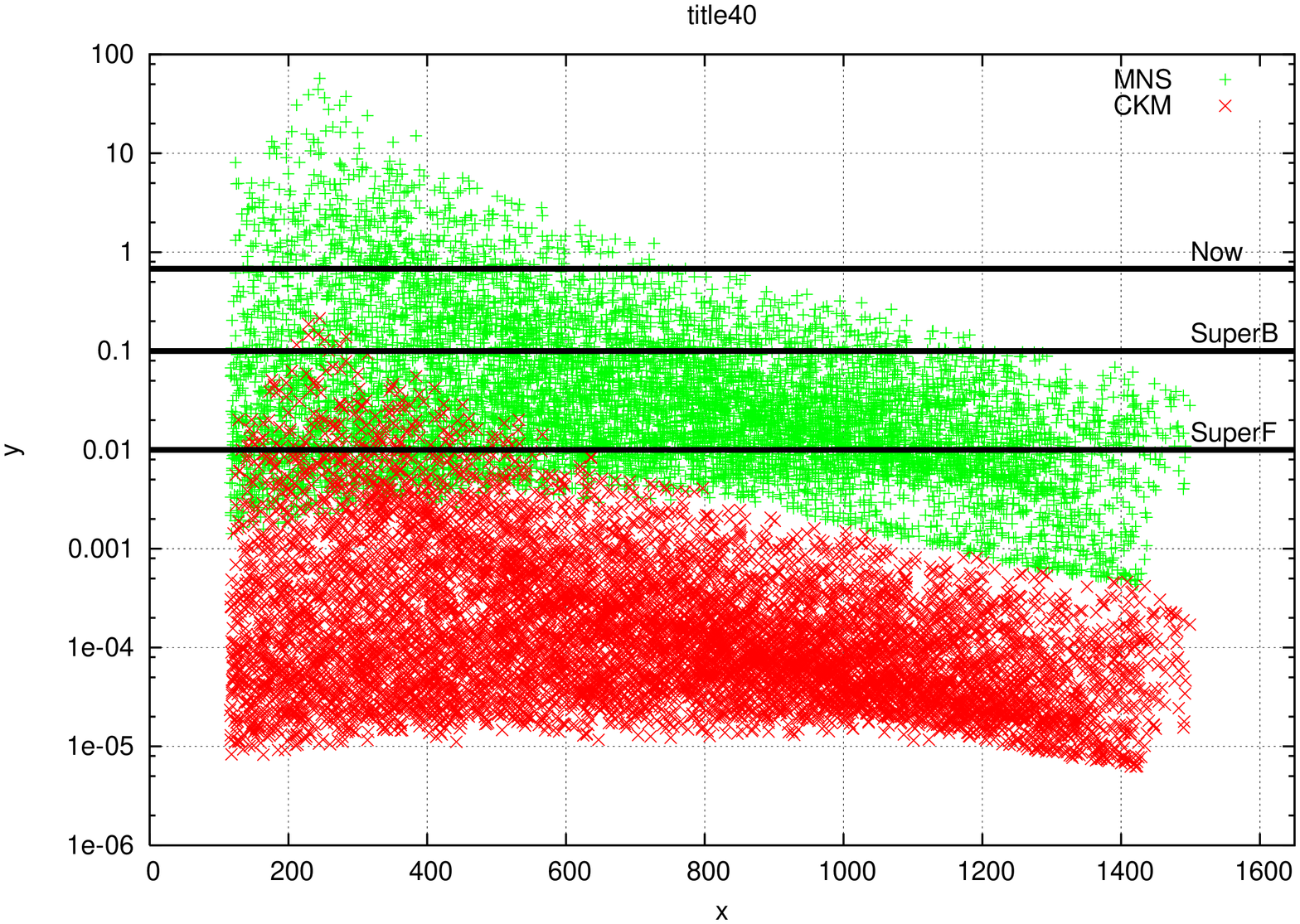}
\caption{\label{tmgplot}Scaled BR($\tau\to \mu\,\gamma$) vs. $M_{1/2}$.
The plots are obtained by scanning the LHC accessible SUSY--GUT
parameter space at
fixed $\tan\beta$.
 The horizontal
lines are the present (B factories), future (SuperKEKB) and
planned (Super Flavour factory) experimental sensitivities.}
\end{figure}

\begin{figure}[h]
\begin{center}
\psfrag{x}[c]{\huge{$m_0$ (GeV)}}
\psfrag{y}[c]{\huge{$M_{1/2}$ (GeV)}}
\psfrag{tmgCKM40}[c]{\huge{CKM case, $\tan\beta=40$}}
\psfrag{tmgMNS10}[c]{\huge{PMNS case, $\tan\beta=10$}}
\psfrag{tmug1E-10}[c]{\Large{$BR=10^{-10}$}}
\psfrag{tmug1E-9}[c]{\Large{$BR=10^{-9}$}}
\psfrag{tmug1E-8}[c]{\Large{$BR=10^{-8}$}}
\includegraphics[angle=-90, width=0.48\textwidth]{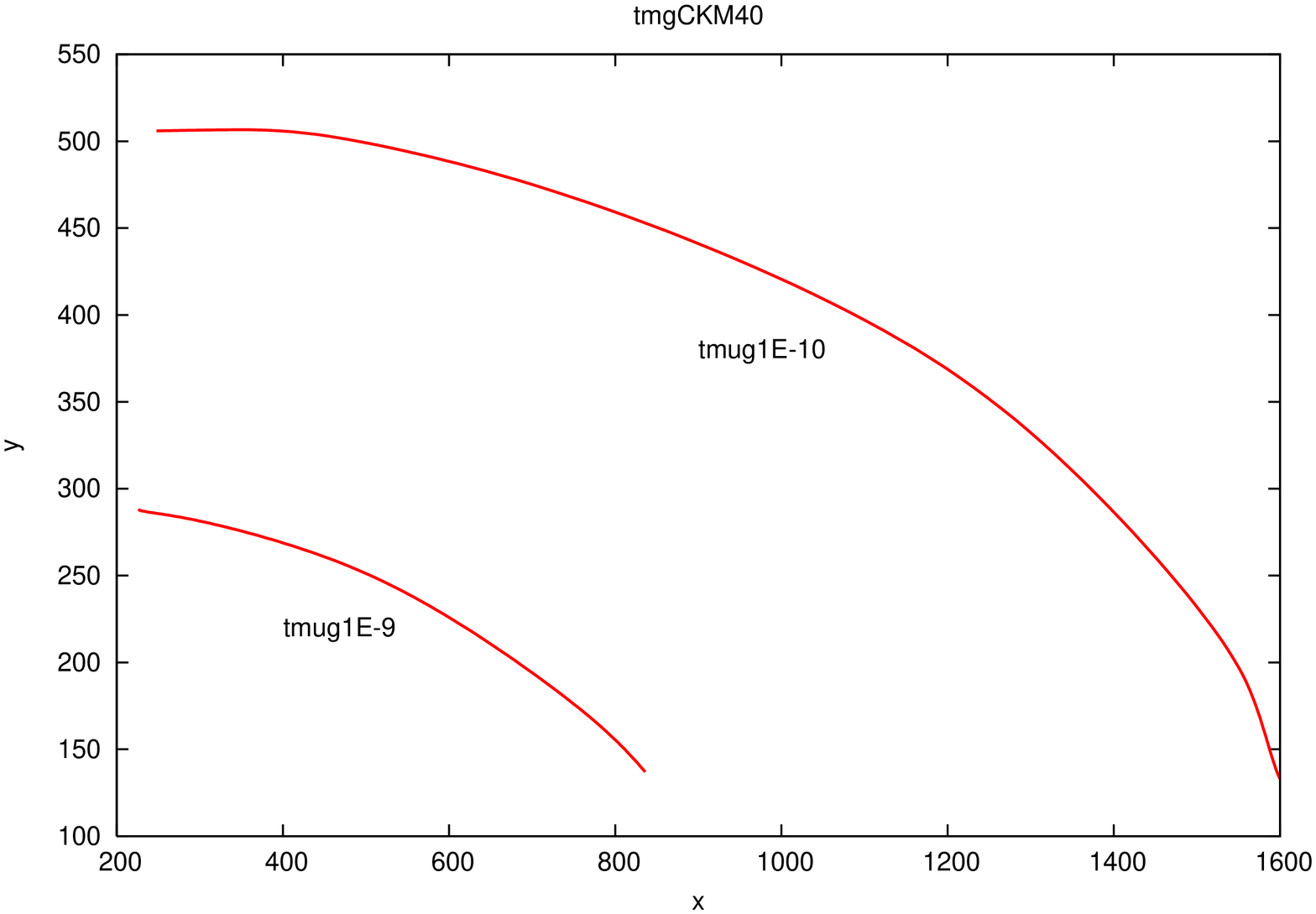}
\includegraphics[angle=-90, width=0.48\textwidth]{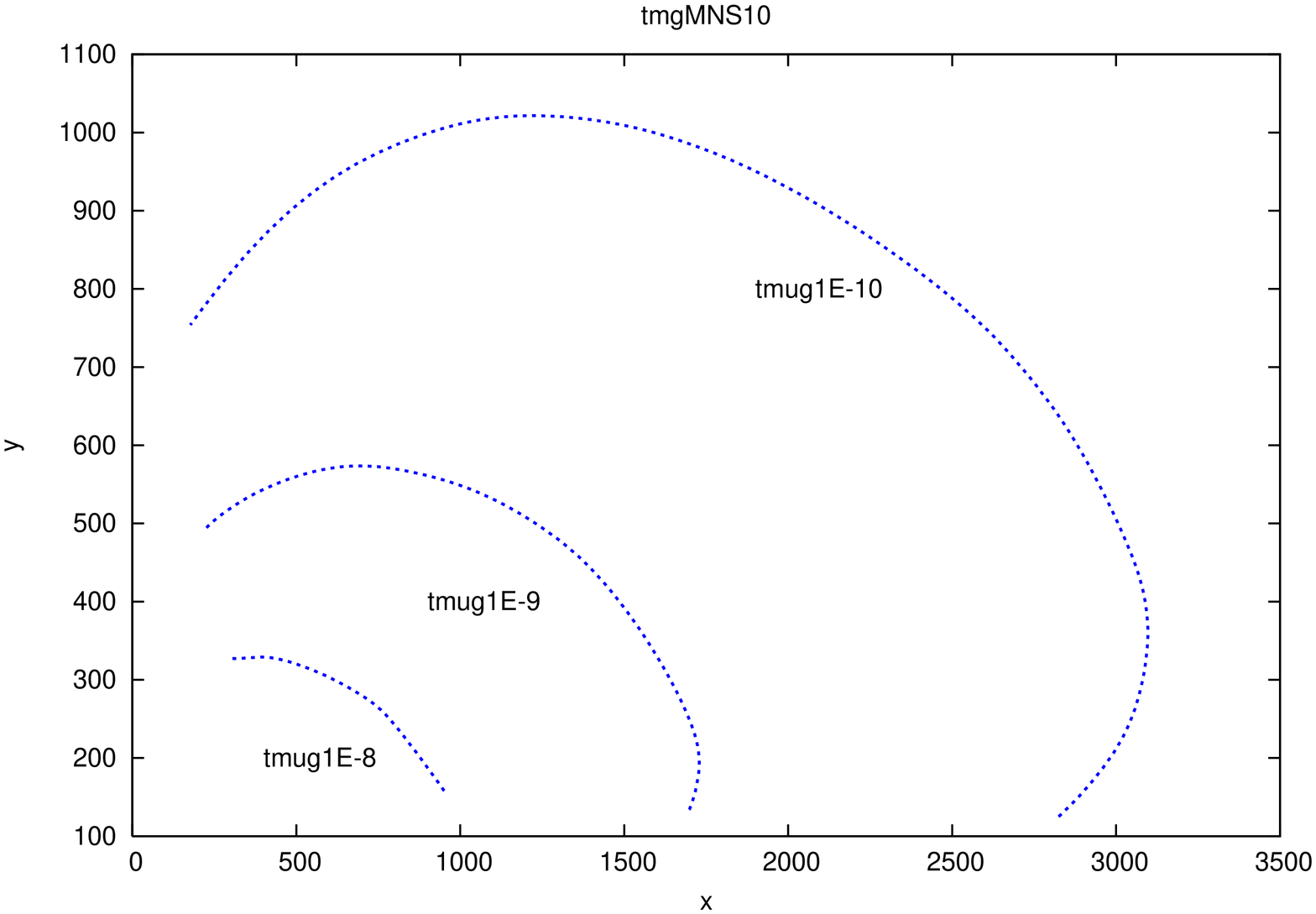}
\end{center}
\caption{\label{tmgcontour}Contour plots at fixed $BR(\tau \to \mu\,\gamma)$
in the $(m_0, M_{1/2})$ plane, at $A_0=0$ in the CKM high $\tan\beta$
case and in the PMNS $t_\beta=10$ one. 
Note that while the planes are presently unconstrained, the Super Flavour 
factory
sensitivity of $\mathcal{O}(10^{-9})$
 would be able to probe  much of the PMNS case at low $t_\beta$ 
and the $(m_0,m_{\tilde{g}})<900$ GeV portion of the high
$\tan\beta$ CKM case.}
\end{figure}

As far as Beauty factories \cite{belletmg, babar, belletalk}
are concerned, we see 
from Fig. \ref{tmgplot}, that even with the present bound
it is possible to rule out part of the PMNS high $\tan\beta$ regime; the
planned accuracy of the SuperKEKB \cite{superKEKB} machine 
$\sim \mathcal{O}(10^{-8})$  will allow to test much 
of high $\tan\beta$ region  and will start probing the low $\tan\beta$
PMNS case, with a sensitivity to soft masses as high as
$(m_0, m_{\tilde{g}}) \lesssim 900$ GeV.
The situation changes dramatically if one takes into account the 
possibility of a Super Flavour factory (Fig. \ref{tmgplot}, \ref{tmgcontour}):
taking the sensitivity of the most promising
$\tau \to \mu \, \gamma$ process to $\sim\mathcal{O}(10^{-9})$, the
PMNS case will be nearly ruled out in the high $\tan\beta$ regime
and severely constrained in the low $\tan\beta$ one; as for
the CKM case it would be tested in the 
$(m_0, m_{\tilde{g}}) \lesssim 900$ GeV region, provided that
$\tan\beta$ is high. 

The conclusions (Table \ref{tmgconc})
are that with the planned improvements of the KEK facility the
$U_{e3}$ independent $\tau \to \mu \, \gamma$ process will
allow us to test much of the PMNS scenario.
A Super Flavour factory would much improve the situation, as it
would be able to almost completely probe the PMNS case and to test
the minimal mixing, high $\tan\beta$ scenario up to soft masses of 600 GeV.

\begin{table}[h]
\caption{\label{tmgconc}Reach in $(m_0, m_{\tilde{g}})$ of the 
present and planned experiment from
their $\tau \to \mu \, \gamma$ sensitivity.}
\begin{ruledtabular}
\begin{tabular}{lcccc}
&\multicolumn{2}{c}{PMNS}&\multicolumn{2}{c}{CKM}\\
Exp.
&$t_\beta=40$ & $t_\beta=10$ &$t_\beta=40$ & $t_\beta=10$
\\ \hline 
BaBar, Belle & 
 1.2 TeV  & no & no & no
\\
SuperKEKB &
 2 TeV & 0.9 TeV & no & no
\\
Super Flavour \footnote{Post--LHC era proposed/discussed experiment}&
 2.8 TeV & 1.5 TeV  &0.9 TeV  & no
\end{tabular}
\end{ruledtabular}
\end{table}

\subsection{Probing the PMNS case with $U_{e3}\approx0$ at MEG}

We have seen that if a Super Flavour factory will be built, the
 $\tau \to \mu \,\gamma$ process will be highly complementary to
the $\mu \to e \, \gamma$ one as a probe of SUSY--GUT scenarios,
 with the added bonus of being
$U_{e3}$ independent. As a Super Flavour factory is just a proposed
experiment, whereas MEG will sure be operating, it is nevertheless
interesting to ask what is the probing capability of such an experiment
in the PMNS case, if $U_{e3}$ happens to be vanishing small, or even 0.

In the case that $U_{e3}=0$ equation \eref{leadingloght}  is no more
a good approximation to the running of the off-diagonal LL entries, as
we have to resort to the 2nd generation entries:
\begin{eqnarray}
\label{mns1b}
(\delta_{LL})_{\mu e}&=& -{3  \over 8 \pi^2}~
Y_c^2 U_{e2} U_{\mu2} \ln{M_{X} \over M_{R_2}} .
\end{eqnarray} 
Here the off-diagonal contribution
in slepton masses, now being proportional to
the square of the charm Yukawa 
$Y_c$ are much smaller, 
in fact even  smaller than the CKM contribution by a factor 
\begin{equation}{Y_c^2~ U_{\mu 2} ~U_{e 2} 
\ln(M_{X}/M_{R_2}) \over  Y_t^2 ~V_{td} ~V_{ts} 
\ln(M_{X}/M_{R_3})}
\sim  \mathcal{O}(10^{-2}).
\label{suppression}
\end{equation}
The point is that the estimate \eref{suppression} misses
and important point.
The PMNS case is the case where the $R$ matrix
is the identity; but we should keep in mind at what scale we
should enforce this. Because the angle $U_{e3}$ runs with 
the energy scale and $U_{e3}\approx 0$ at the weak scale
does not necessarily mean $U_{e3} \approx 0$ at high scale. 
Even for hierarchical spectra,
where the running effects are small,  the induced RG effects 
in the soft spectrum could be large, leading to large enough
$\mu \to e \, \gamma$. The running effect of the neutrino 
mixing angle can be estimated by using the
neutrino RG \cite{chankowski-pokorski, lindnerantusch} equations.

Moreover, as we have seen in section III, in a SUSY--GUT framework
we have also sizable subleading contribution to the amplitude
of the $\mu \to e \, \gamma$ process coming form the 
$(\delta_{RR})_{e\mu}$
insertion and from the double insertions 
$(\delta_{RR})_{e\tau} (\delta_{LL})_{\tau\mu}$; the interplay between
the RG enhancement of $U_{e3}$ and the amplitudes coming
from the subleading insertions will be thoroughly discussed
in an upcoming publication \cite{noiUe3}. 
 
\begin{figure}[h]
\psfrag{y}[c]{\huge{$BR(\mu \to e\,\gamma)\cdot10^{11}$ }}
\psfrag{x}[c]{\huge{$M_{1/2}$ (GeV)}}
\psfrag{title10}[c]{\huge{Comparison of $\mu\to e\,\gamma$ at $\tan\beta=10$ in
different scenarios}}
\psfrag{title40}[c]{\huge{Comparison of $\mu\to e\,\gamma$ at $\tan\beta=40$ in
different scenarios}}
\psfrag{CKM}[r]{\Large{CKM}}
\psfrag{007}[r]{\Large{PMNS $U_{e3}=0.07$}}
\psfrag{000}[r]{\Large{PMNS $U_{e3}=0$}}
\includegraphics[angle=-90, width=0.48\textwidth]{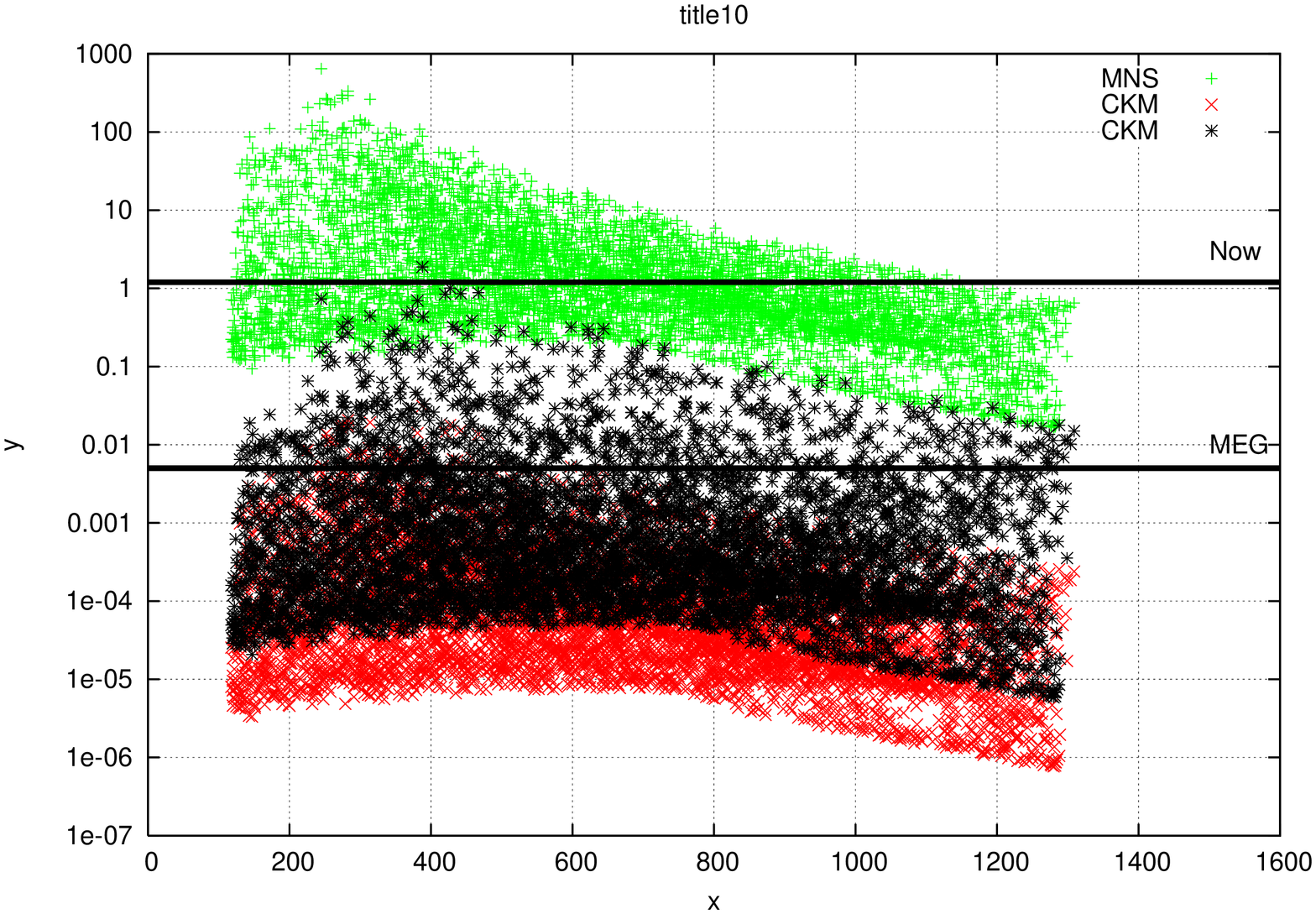}
\includegraphics[angle=-90, width=0.48\textwidth]{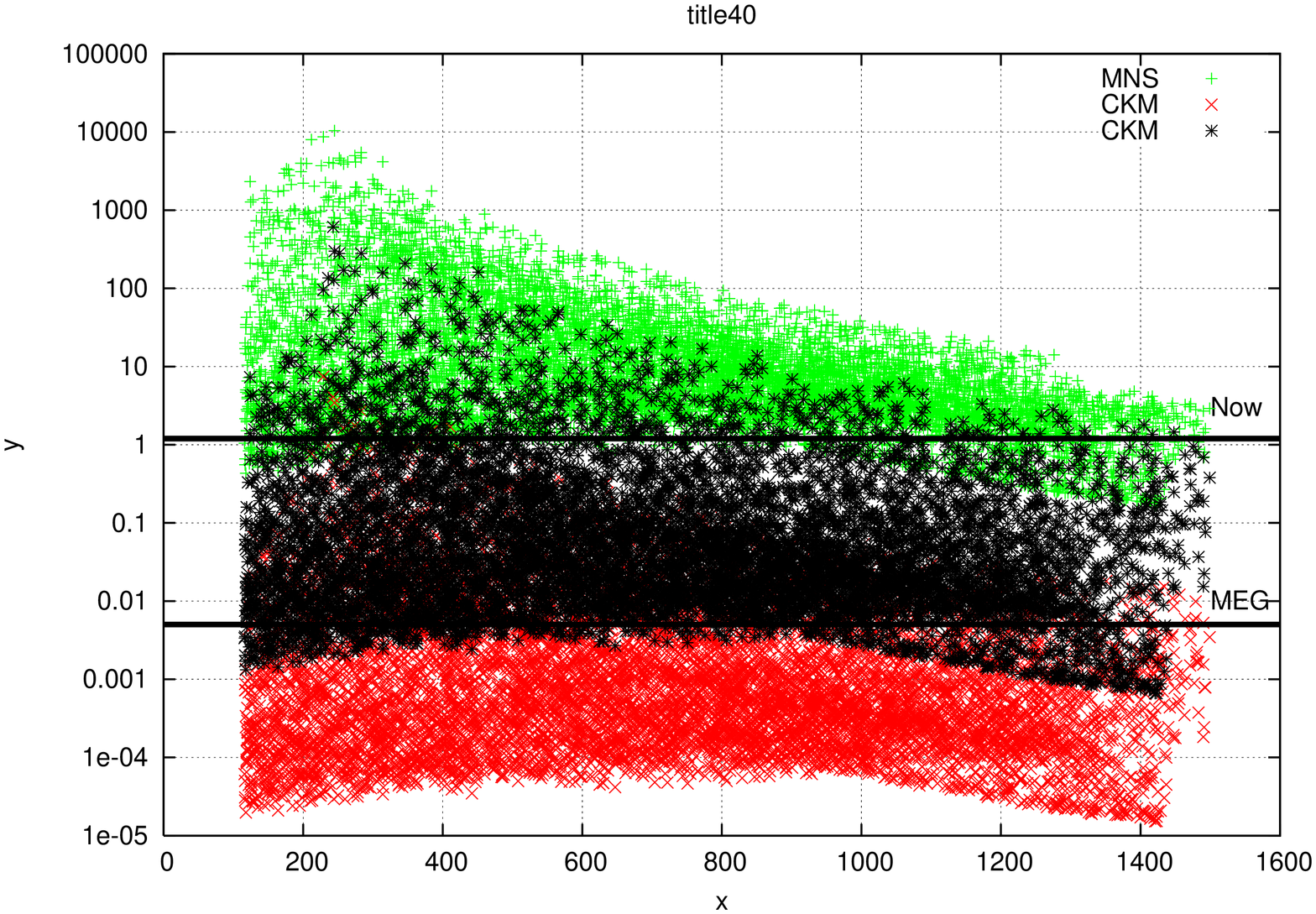}
\caption{\label{ue3plot} BR$(\mu\to e\,\gamma)$ as a probe of 
different SUSY--GUT
scenarios. The plot are obtained by scanning the LHC accessible
parameter space at fixed $\tan\beta$. The lines are the present 
(MEGA) and future (MEG) experimental sensitivities.
We see that MEG will completely test the PMNS scenario for
$U_{e3}$ close to the CHOOZ bound and severely constrain it for $U_{e3}=0$.}
\end{figure}

The results for the PMNS mixing with $U_{e3} = 0$ (defined at the
weak scale) are shown in Fig. \ref{ue3plot}. We see that even for 
low $\tan\beta$ the branching
ratio is never lower than that of the CKM case, giving
a proof that the CKM case is really a representative of a `minimal
mixing' case.
We see (Table \ref{ue3conc})
that, given the present experimental LFV rates bounds, the $U_{e3} = 0$ 
PMNS case is better constrained by
the $\tau \to \mu \, \gamma$ than by the $\mu \to e \, \gamma$
process.
MEG will be able to probe much of this scenario:
for high values of $\tan\beta$ almost all the LHC accessible
 parameter space will be probed, whereas if $\tan\beta$ happens to be
small it will be probed up to 
$(m_0, m_{\tilde{g}}) \lesssim 1100$ GeV. 
We thus can state that if $\tan\beta$ is high the MEG experiment
will probe the PMNS case better than the $\tau\mu$ sector
experiments, irrespectively of the value of $U_{e3}$, and
with an accuracy comparable to that of the SuperKEKB machine
if $\tan\beta$ is small. On the
other hand, a Super Flavour factory would for sure supersede MEG.

\begin{table}
\caption{\label{ue3conc}Reach in $(m_0, m_{\tilde{g}})$ of the 
past and upcoming experiments from
their $\mu \to e\,\gamma$ sensitivity.  LHC means that all the LHC testable
parameter space will be probed.}
\begin{ruledtabular}
\begin{tabular}{lcc}
&\multicolumn{2}{c}{PMNS, $U_{e3}=0$}\\
Exp. & $t_\beta=40$ &$t_\beta=10$ \\
\hline
MEGA & 1.1 TeV  & no \\
MEG & LHC & 1.1 TeV 
\end{tabular}
\end{ruledtabular}
\end{table}

\subsection{The PRISM/PRIME experiment at J-PARC}

Since the experimental signal is very well defined,
the $\mu \to e$ conversion in Nuclei poses very good prospects
as a probe of lepton flavour violating scenarios. In SUSY--GUT frameworks
the main contribution to the amplitude comes from the
penguin diagram that is also responsible for the 
FV $\mu \to e \, \gamma$ amplitude. There is thus a strong
correlation between these two processes, the $\mu \to e$
conversion being suppressed by a factor $\sim Z\alpha/\pi$ 
with respect to the flavour violating decay $\mu \to e \, \gamma$.

The present bounds on $\mu \to e$ conversion come from
the SINDRUM II experiment at PSI, that gave bounds on
conversion rates in different Nuclei. For instance, the 
bound for the conversion in Titanium ($4.3 \cdot 10^{-12}$)
is almost as good as the current MEGA bound on $\mu \to e \,\gamma$ 
($1.1 \cdot 10^{-11}$)
in constraining the
SUSY--GUT parameter space, but it will be superseded by the 
future MEG sensitivity. To achieve a sensitivity to SUSY--GUTs
scenarios that is comparable to the MEG experiment, a 
$\mu \to e$ conversion experiment in Titanium would
need a sensitivity of $\mathcal{O}(10^{-15})$.
This would require an high intensity muon source and
an experimental apparatus that provides
a very good resolution in the energy of the emitted 
electron, to discriminate with high accuracy the
$\mu \to e$ conversion versus the $\mu$ decay in orbit.
The J-PARC experiment PRISM/PRIME 
\cite{prismprime:loi} addresses these issues
by means of an innovative $\mu$ source
(Phase Rotated Intense Slow Muons, PRISM),
with an intensity of $10^{11}-10^{12}$ $\mu/s$, 
and its $\mu \to e$ conversion in Ti dedicated experiment
(PRIME: PRISM $\mu-e$ conversion experiment); 
the planned sensitivity of the experiment
is of $4 \cdot 10^{-18}$, with the possibility of
improving it by upgrading the PRISM machine
intensity to $10^{14}$ $\mu/s$.

\begin{figure}[h]
\psfrag{y}[c]{\huge{$CR(\mu \to e)\cdot10^{12}$ in Ti}}
\psfrag{x}[c]{\huge{$M_{1/2}$ (GeV)}}
\psfrag{title10}[c]{\huge{$\mu\to e$ in Ti at $\tan\beta=10$}}
\psfrag{title40}[c]{\huge{$\mu\to e$ in Ti at $\tan\beta=40$}}
\psfrag{CKM}[r]{\Large{CKM}}
\psfrag{MNS}[r]{\Large{PMNS}}
\includegraphics[angle=-90, width=0.48\textwidth]{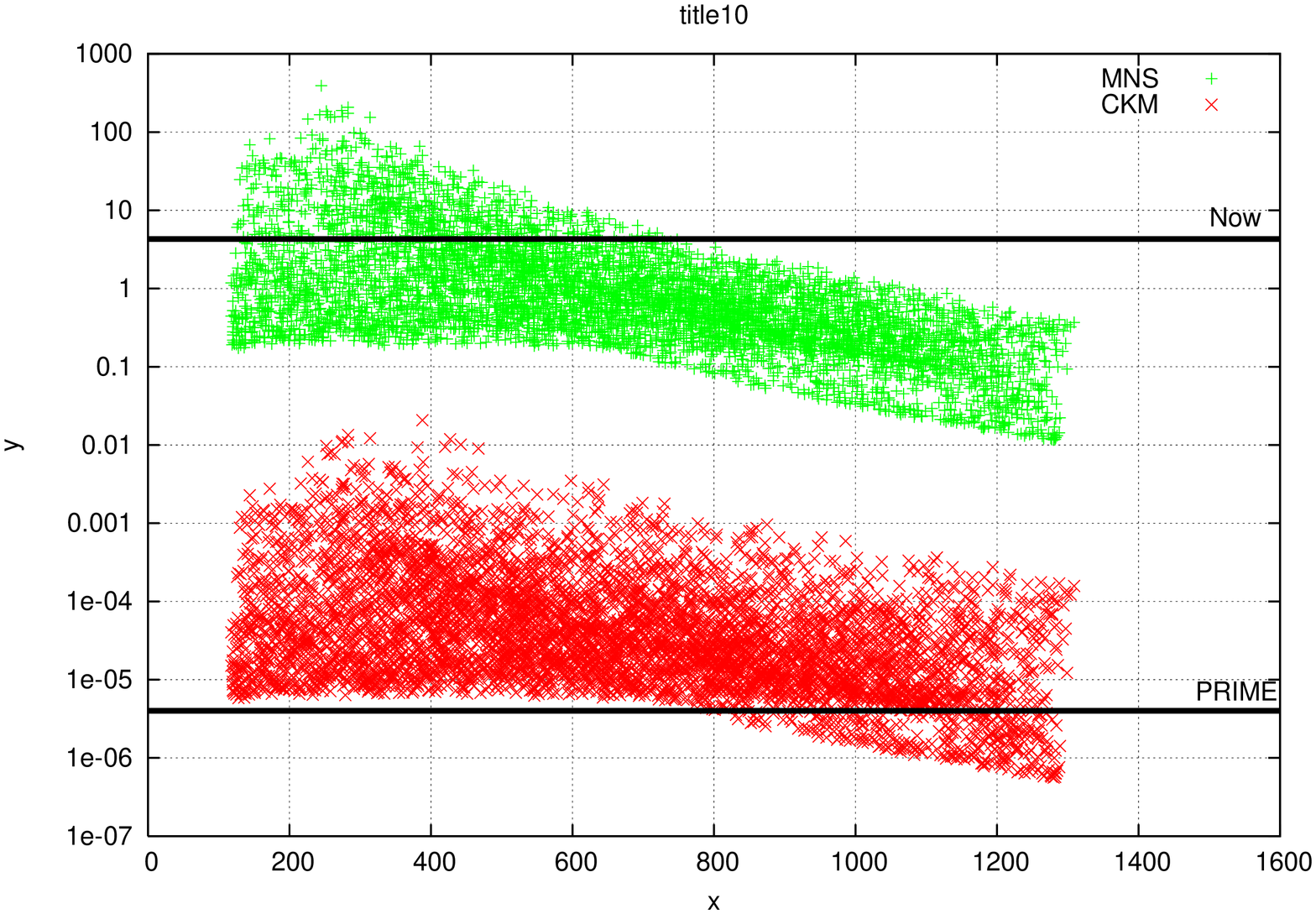}
\includegraphics[angle=-90, width=0.48\textwidth]{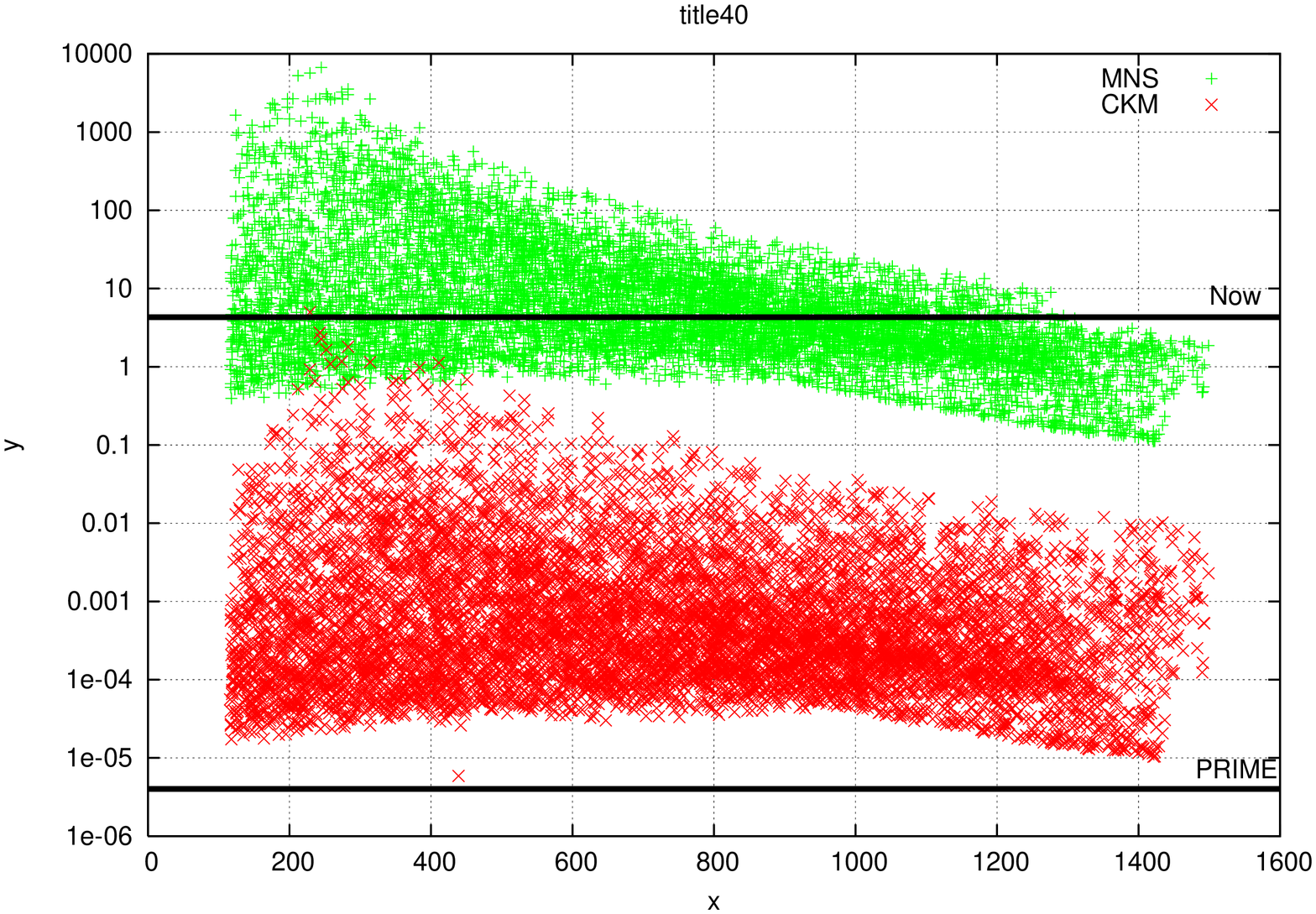}
\caption{\label{conv:scatter} $\mu \to e$ in Ti as a probe of
SUSY--GUT scenarios. The plots are obtained by scanning the LHC
accessible parameter space. 
The horizontal lines are the present
(SINDRUM II) bound and the planned (PRISM/PRIME) sensitivity
to the process. We see that PRIME would be able to
severely constrain the low $\tan\beta$, low mixing
angles case and to completely test the other scenarios.}
\end{figure}

\begin{figure}[h]
\psfrag{x}[c]{\huge{$m_0$ (GeV)}}
\psfrag{y}[c]{\huge{$M_{1/2}$ (GeV)}}
\psfrag{ckm10}[c]{\huge{$\mu\to e$ in Ti at $\tan\beta=10$, CKM case}}
\psfrag{ckm40}[c]{\huge{$\mu\to e$ in Ti at $\tan\beta=40$, CKM case}}
\psfrag{10E-14}[l]{\Large{$CR=10^{-14}$}}
\psfrag{10E-15}[l]{\Large{$CR=10^{-15}$}}
\psfrag{10E-16}[l]{\Large{$CR=10^{-16}$}}
\psfrag{10E-17}[l]{\Large{$CR=10^{-17}$}}
\psfrag{10E-18}[l]{\Large{$CR=10^{-18}$}}
\includegraphics[angle=-90, width=0.48\textwidth]{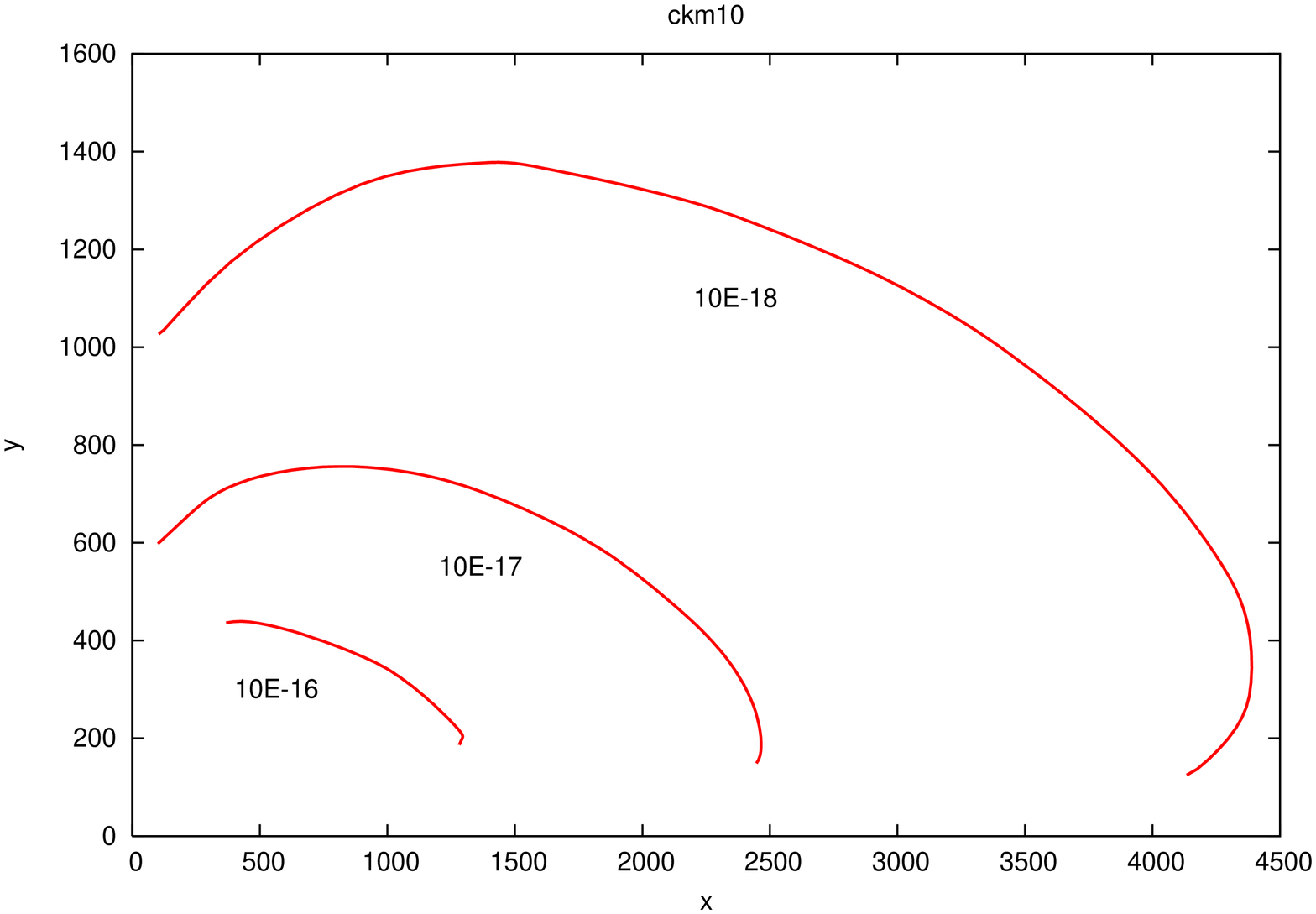}
\includegraphics[angle=-90, width=0.48\textwidth]{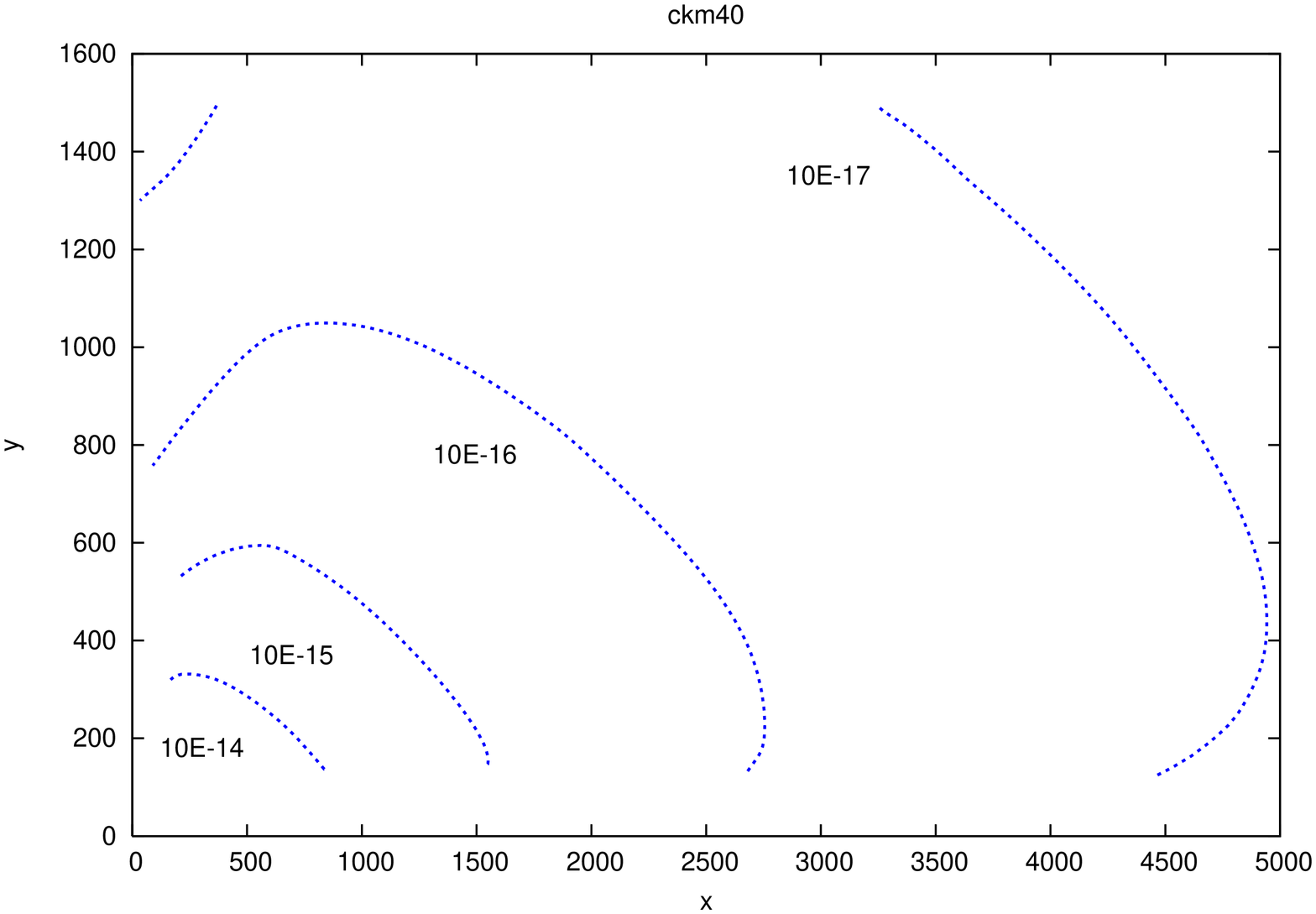}
\caption{\label{conv:contour} 
Contour plots at $A_0=0$ of the parameter space region within
reach of different $\mu \to e$ in Ti CR sensitivities in the
CKM case for low and high $\tan\beta$. We see that the PRIME
experiment will be able to test the CKM $t_\beta=10$ case
for $(m_0, m_{\tilde{g}}) \lesssim 2800$ GeV and the $t_\beta=40$
even beyond LHC reach.}
\end{figure}

Although the experiment has not yet been approved, the construction
of the PRISM machine has already begun and should be 
completed in five years \cite{prismprime:kuno}. 
It is thus timely to ask
what will be the power of the post--LHC PRIME experiment to
discriminate between the different SUSY--GUT scenarios
in the case that the LHC finds evidence for SUSY.
As can be seen from Fig. \ref{conv:scatter} and \ref{conv:contour}
the PRIME experiment would be able to really test our
SUSY--GUT ansatz (Table \ref{conv:table}): the high $\tan\beta$ case would
be tested in both the large and small mixing angles 
scenarios, even beyond the reach of the LHC. As
for the low $\tan\beta$ scenario, the PMNS case
would be completely tested and much of the CKM case
would be within reach: masses as high as 
 $(m_0, m_{\tilde{g}}) \lesssim 2800$ GeV could be
probed.

\begin{table}[h]
\caption{\label{conv:table}Reach in $(m_0, m_{\tilde{g}})$ of the
present and planned experiment from
their $\mu \to e$ conversion sensitivity.
LHC means that all the LHC testable
parameter space will be probed. All means that masses
as high as $(m_0, m_{\tilde{g}}) \lesssim 5$ TeV will be probed.}
\begin{ruledtabular}
\begin{tabular}{lcccc}
&\multicolumn{2}{c}{PMNS}&\multicolumn{2}{c}{CKM}\\
Exp.
&$t_\beta=40$ & $t_\beta=10$ &$t_\beta=40$ & $t_\beta=10$
\\ \hline
SINDRUM II &
 2 TeV  & 1.3 TeV & no & no
\\
MECO\footnote{MECO \cite{meco} was terminated by the NSF on Fall 2004. 
The values are given as a reference comparison} &
 all & all & 2.6 TeV & 1.3 TeV
\\
PRISM/PRIME\footnote{Post--LHC era proposed experiment}&
 all & all  & LHC  & 2.8 TeV
\end{tabular}
\end{ruledtabular}
\end{table}

As the PRIME experiment would be a post--LHC era
experiment its capability of testing and ruling out
so many different SUSY--GUT scenarios is most
interesting. It would be an ideal complement
to the findings of the LHC in the case that it
gets positive evidence for low
energy supersymmetry.

\section{LFV rates at SPS benchmark points}

In this section we  discuss the possibility of detecting
supersymmetry at the SPS benchmark points \cite{sps} by means
of LFV experiments. We  concentrate on the SPS points defined
for mSUGRA/CMSSM framework. These take in to consideration various
constraints, including relic density requirements, in addition to 
what we have considered here. We note that some of
these points will be ruled out in the light of new WMAP data if one
requires a purely Bino dark matter. As of now, there is no corresponding
definition of SPS points within SUSY--GUTs. In the present work, we consider
the input values of the mSUGRA SPS points in our $SO(10)$ model and study
the impact of flavour violation for that spectra \footnote{In some points,
we notice the need for modifying these numbers within a SUSY--GUT framework.
For example, in {\bf SPS 3}, the LSP and $\tilde{\tau}_1$ are no longer
degenerate, whereas {\bf SPS 4} and {\bf SPS 5} are already in conflict
with experimental measurements.}
We note that for all the points, the PMNS
framework is ruled out by the present MEGA bound on
$\mu\to e \, \gamma$. Furthermore, the PRISM/PRIME
experiment would be able to test all the scenarios.

\begingroup
\squeezetable
\begin{table*}[t]
\caption{\label{sps1asps1b}LFV rates for points {\bf SPS 1a} and {\bf SPS 1b}
in the CKM case and in the $U_{e3}=0$ PMNS case. The processes
that are within reach of the future experiments (MEG, SuperKEKB) have been
highlighted in boldface. Those within reach of post--LHC era planned/discussed
experiments (PRISM/PRIME, Super Flavour factory) highlighted in italics.}
\begin{ruledtabular}
\begin{tabular}{lccccccccc}
&\multicolumn{2}{c}{{\bf SPS 1a}}
&\multicolumn{2}{c}{{\bf SPS 1b}}
&\multicolumn{2}{c}{{\bf SPS 2}}
&\multicolumn{2}{c}{{\bf SPS 3}}
& Future \\
Process & CKM & $U_{e3}=0$ & CKM & $U_{e3}=0$ &
 CKM & $U_{e3}=0$ & CKM & $U_{e3}=0$ 
&Sensitivity \\
\hline 
BR($\mu \to e\,\gamma$) & ${\bf 3.2 \cdot 10^{-14}}$ 
                        & ${\bf 3.8 \cdot 10^{-13}} $
                        & ${\bf 4.0 \cdot 10^{-13}}$
                        & ${\bf 1.2 \cdot 10^{-12} }$
			& $1.3 \cdot 10^{-15}$ 
                        & $8.6 \cdot 10^{-15} $
                        & $1.4 \cdot 10^{-15}$
                        & $ 1.2 \cdot 10^{-14} $
                        & $\mathcal{O}(10^{-14})$ \\ 
BR($\mu \to e\,e\,e$ )  & $2.3 \cdot 10^{-16}$ 
                        & $2.7 \cdot 10^{-15}$
                        & $2.9 \cdot 10^{-16}$
                        & $8.6 \cdot 10^{-15}$
			& $9.4 \cdot 10^{-18}$ 
                        & $6.2 \cdot 10^{-17}$
                        & $1.0 \cdot 10^{-17}$
                        & $8.9 \cdot 10^{-17}$
                        & $\mathcal{O}(10^{-14})$ \\ 
CR($\mu \to e$ in Ti) & ${\it 2.0 \cdot 10^{-15} }$  
                          & ${\it 2.4 \cdot 10^{-14} }$
                          & ${\it 2.6 \cdot 10^{-15} }$
                          & ${\it 7.6 \cdot 10^{-14}}$
			  & ${\it 1.0 \cdot 10^{-16} }$  
                          & ${\it 6.7 \cdot 10^{-16} }$
                          & ${\it 1.0 \cdot 10^{-16} }$
                          & ${\it 8.4 \cdot 10^{-16} }$
                          & $\mathcal{O}(10^{-18})$ 
\vspace{0.75mm}\\
BR($\tau \to e\,\gamma$) & $2.3 \cdot 10^{-12}$  
                         & $6.0 \cdot 10^{-13}$
                         & $3.5 \cdot 10^{-12}$
                         & $1.7 \cdot 10^{-12}$
			& $1.4 \cdot 10^{-13}$  
                         & $4.8 \cdot 10^{-15}$
                         & $1.2 \cdot 10^{-13}$
                         & $4.1 \cdot 10^{-14}$
                         & $\mathcal{O}(10^{-8}) $ \\ 
BR($\tau \to e\,e\,e$) & $2.7 \cdot 10^{-14}$  
                       & $7.1 \cdot 10^{-15}$
                       & $4.2 \cdot 10^{-14}$
                       & $2.0 \cdot 10^{-14}$
			& $1.7 \cdot 10^{-15}$  
                       & $5.7 \cdot 10^{-17}$
                       & $1.5 \cdot 10^{-15}$
                       & $4.9 \cdot 10^{-16}$
                       & $\mathcal{O}(10^{-8}) $ \\ 
BR($\tau \to \mu\,\gamma$) & $5.0 \cdot 10^{-11}$  
                           & ${\bf 1.1 \cdot 10^{-8}}$
                           & $7.3 \cdot 10^{-11}$
                           & ${\bf 1.3 \cdot 10^{-8} }$
			& $2.9 \cdot 10^{-12}$  
                           & ${\it 7.8 \cdot 10^{-10}}$
                           & $2.7 \cdot 10^{-12}$
                           & ${\it 6.0 \cdot 10^{-10} }$
                           &$\mathcal{O}(10^{-9}) $ \\ 
BR($\tau \to \mu\, \mu\, \mu$) & $1.6 \cdot 10^{-13}$  
                               & $3.4 \cdot 10^{-11}$
                               & $2.2 \cdot 10^{-13}$
                               & $3.9 \cdot 10^{-11}$
				& $8.9 \cdot 10^{-15}$  
                               & $2.4 \cdot 10^{-12}$
                               & $8.7 \cdot 10^{-15}$
                               & $1.9 \cdot 10^{-12}$
                               & $\mathcal{O}(10^{-8}) $ \\ 
\end{tabular}
\end{ruledtabular}
\end{table*}
\endgroup

The `typical' mSUGRA scenario is represented by SPS points
{\bf 1a} at low $\tan\beta$ and {\bf 1b} at relatively high
$t_\beta$
\begin{eqnarray*}
\textrm{\bf SPS 1a} &:&
m_0 =100, \: 
M_{1/2} = 250, \:
A_0 = - 100, \:
t_\beta = 10 \\
&&
m_h = 112, \: 
m_{\tilde{t}} = 375, \: 
m_{\tilde{g}} = 612 \\
\textrm{\bf SPS 1b} &:&
m_0 = 200, \: 
M_{1/2} = 400, \: 
A_0 = 0, 
t_\beta = 30\\
&&
m_h = 120, \: 
m_{\tilde{t}} = 636 \: 
m_{\tilde{g}} = 980 
\end{eqnarray*}
where the values are given in GeV and we have also given 
 the values of three low energy observable
($m_h$, $m_{\tilde{t}}$, $m_{\tilde{g}}$) as obtained from the
 routine \footnote{All the SPS points have $\mu>0$.}. 
We see that
point {\bf 1a} is already ruled out by the bound on the lightest
Higgs mass. We are including it as it lays at the boundary
of the experimentally ruled out region,
 so that a further improved version of our code 
  could
give the small correction that is needed to satisfy the present bound.
The CKM scenario and the PMNS case at
$U_{e3}=0$ are unscathed by the present bounds.
We see (Table \ref{sps1asps1b})
that the PMNS $U_{e3}=0$ scenario
will be within reach of both MEG and SuperKEKB, for
the two benchmark points, while the CKM case could escape
MEG detection, as the predicted BR for both points
are at the boundary of the planned sensitivity.

The {\bf SPS 2} benchmark point lies in the so-called `focus
point' region \cite{feng}
\begin{eqnarray*}
\textrm{\bf SPS 2} &:&
m_0 =1450\textrm{, }
M_{1/2} = 300 \textrm{, }
A_0 = 0 \textrm{, }
t_\beta = 10 \\
&&
m_h = 124 \textrm{, }
m_{\tilde{t}} = 940 \textrm{, }
m_{\tilde{g}} = 735
\end{eqnarray*}
where all the masses are given in GeV.
From Table \ref{sps1asps1b} we see that the PMNS $U_{e3}=0$ 
scenario will be within reach of the proposed Super Flavour factory;
as for the other processes they will escape detection.

The mSUGRA/CMSSM `coannihilation region' \cite{coa} has its representative in 
point {\bf SPS 3}. In this region a rapid coannihilation between
the neutralino LSP and the stau NLSP will give rise to a 
sufficiently low relic abundance: for this reason, we are
also giving $m_{\tilde{\tau}}$ and $m_{LSP}$ as low energy
observables (all masses in GeV) 
\begin{eqnarray*}
\textrm{\bf SPS 3} &:&
m_0 =90 \textrm{, }
M_{1/2} = 400 \textrm{, }
A_0 = 0 \textrm{, }
t_\beta = 10 \\
&&
m_h = 119 \textrm{, }
m_{\tilde{t}} = 631 \textrm{, }
m_{\tilde{g}} = 980 \\
&&
m_{\tilde{\tau}} = 270 \textrm{, }
m_{LSP} = 185
\end{eqnarray*}
This point will be within reach of the proposed Super Flavour
factory (Table \ref{sps1asps1b}) in the PMNS $U_{e3}=0$ scenario.

The mSUGRA scenario at high $\tan\beta$ has it benchmark in point
{\bf SPS 4}
\begin{eqnarray*}
\textrm{\bf SPS 4} &:&
m_0 =400 \textrm{, }
M_{1/2} = 300 \textrm{, }
A_0 = 0 \textrm{, }
t_\beta = 50 
\end{eqnarray*}
where all masses are in GeV. This point is ruled out, because it 
gives a non-viable vacuum.

The point {\bf SPS 5}, that corresponds to a scenario of relatively
light stop, is ruled out because it predicts a too light Higgs boson
\begin{eqnarray*}
\textrm{\bf SPS 5} &:&
m_0 =150 \textrm{, }
M_{1/2} = 300 \textrm{, }
A_0 = -1000 \textrm{, }
t_\beta = 5 \\
&&
m_h = 102 \textrm{,; }
m_{\tilde{t}} = 275 \textrm{, }
m_{\tilde{g}} = 735 
\end{eqnarray*}
where the dimensional parameters are given in GeV.

As a conclusion (Table \ref{spsconclusion})
we can state that the only scenarios that will for sure
escape detection are the CKM focus point {\bf SPS 2} and
CKM coannihilation region {\bf SPS 3} cases. The {\bf SPS 1a}
and {\bf SPS 1b} CKM scenario are at the boundary of MEG sensitivity
so that probing these scenario, though hard, is nevertheless 
a possibility. The PRISM/PRIME experiment would much improve
the situation, as it would be able to test all the scenarios;
these results would be complemented by that from a Super Flavour
factory.

\begin{table}[h!!!]
\caption{\label{spsconclusion}Capability of past,
 present and future experiment
to detect LFV at the SPS benchmark points. When two experiment
are able to detect the same process, only the less sensitive
experiment is displayed. }
\begin{ruledtabular}
\begin{tabular}{lccc}
Point& CKM & PMNS & PMNS, $U_{e3}=0$ \\
\hline{\bf SPS 1a} 
               	&\begin{tabular}{c}
		MEG (maybe)\\
		PRIME\footnotemark[1]
		\end{tabular}
               	&\begin{tabular}{c}
		MEGA\\
		SINDRUM II\\
		SuperKEKB
		\end{tabular}
		&\begin{tabular}{c}
                MEG\\
		PRIME\footnotemark[1]\\
		SuperKEKB
                \end{tabular}\\
\\{\bf SPS 1b}
		&\begin{tabular}{c}
                MEG (maybe)\\
		PRIME\footnotemark[1]
		\end{tabular}
                &\begin{tabular}{c}
                MEGA\\
                SINDRUM II\\
		SuperKEKB
                \end{tabular}
                &\begin{tabular}{c}
                MEG\\
		PRIME\footnotemark[1]\\
                SuperKEKB
                \end{tabular}\\
\\{\bf SPS 2}
                & PRIME\footnotemark[1] 
                &\begin{tabular}{c}
                MEGA\\
                SINDRUM II\\
		Super Flavour\footnotemark[1]
                \end{tabular}
                &\begin{tabular}{c}
		PRIME\footnotemark[1]\\
                Super Flavour\footnotemark[1]
                \end{tabular}\\
\\{\bf SPS 3}
                &PRIME\footnotemark[1] 
                &\begin{tabular}{c}
                MEGA\\
                SINDRUM II\\
		Super Flavour\footnotemark[1]
                \end{tabular}
                &\begin{tabular}{c}
		PRIME\footnotemark[1] \\
                Super Flavour\footnotemark[1]
                \end{tabular}\\
\end{tabular}
\end{ruledtabular}
\footnotetext[1]{Post--LHC era, planned/discussed experiment}
\end{table}

\section{Conclusions}

In this paper we addressed the capability of past
(MEGA, SINDRUM II), present (BaBar, Belle), upcoming
(MEG, SuperKEKB) and proposed (PRISM/PRIME, Super
Flavour factory) Lepton Flavour Violation experiments
to probe SUSY--GUT scenarios. We have found that these
experiments have strong capabilities to detect SUSY
induced LFV, in some cases even outreaching
the LHC.

\begin{figure}[h]
\psfrag{y}[c]{\huge{$BR(\tau \to \mu\,\gamma)\cdot10^{7}$ }}
\psfrag{x}[c]{\huge{$BR(\mu \to e\,\gamma)\cdot10^{11}$ }}
\psfrag{tmgvsmeg10}[c]{\huge{$\mu\to e\,\gamma$ vs.
$\tau\to\mu\,\gamma$ at  $\tan\beta=10$}}
\psfrag{tmgvsmeg40}[c]{\huge{$\mu\to e\,\gamma$ vs.
$\tau\to\mu\,\gamma$ at  $\tan\beta=40$}}
\psfrag{CKM}[c]{\Large{CKM}}
\psfrag{MNS}[c]{\Large{PMNS $U_{e3}=0.07$}}
\psfrag{UE340}[c]{\Large{PMNS $U_{e3}=0$}}
\psfrag{UE310}[c]{\Large{PMNS $U_{e3}=0$}}
\includegraphics[angle=-90, width=0.48\textwidth]{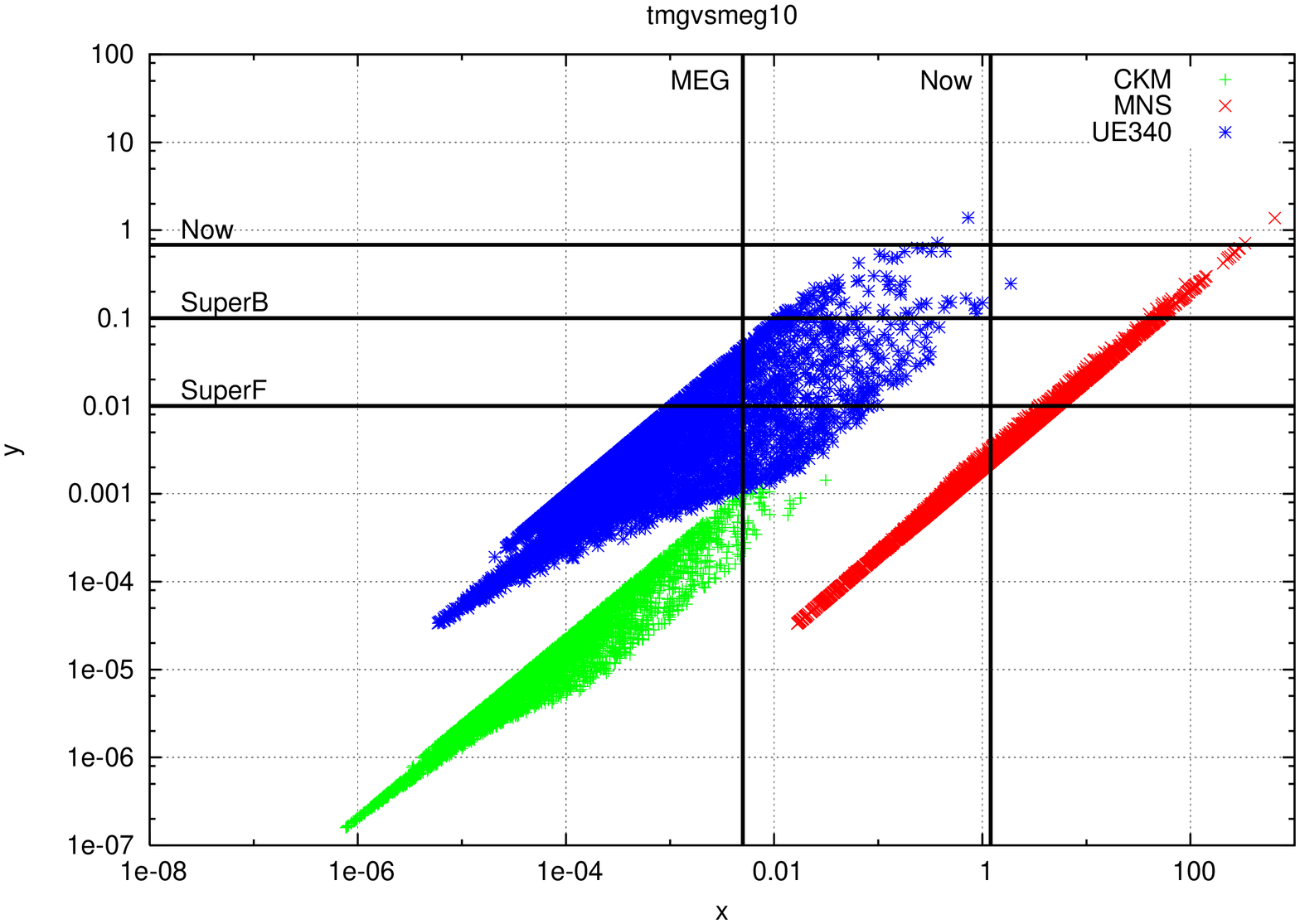}
\includegraphics[angle=-90, width=0.48\textwidth]{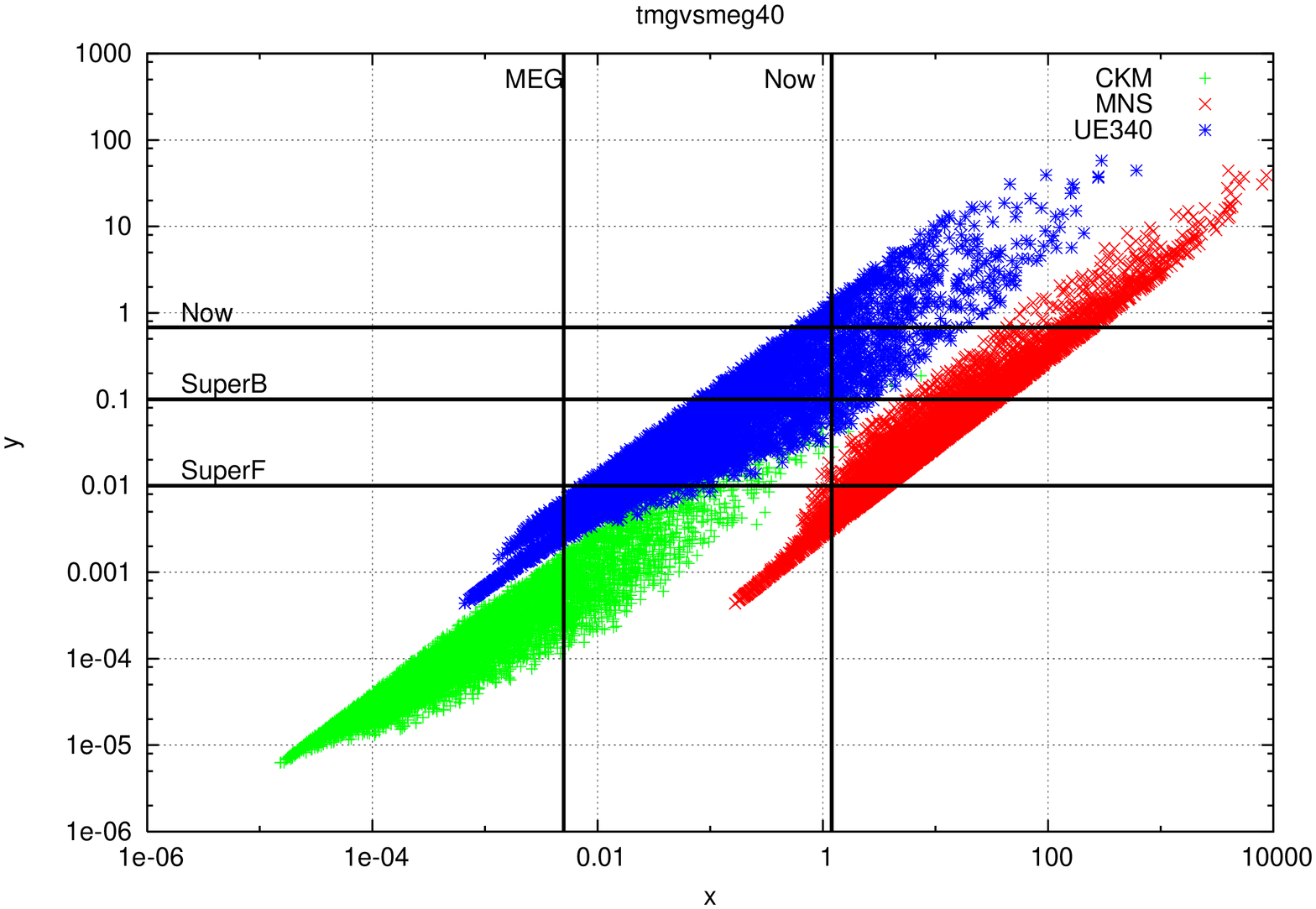}
\caption{\label{ue3confronto}Comparison of  $\mu\to e\,\gamma$ 
and $\tau \to \mu\,\gamma$ as a probes of SUSY--GUTs
scenarios. The plots are done by scanning the LHC accessible
parameter space at fixed $\tan\beta$.
 The lines are the present bounds and future sensitivities.
Let us note that the interplay between MEG and a Super Flavour factory
will leave unscathed only the low $\tan\beta$ CKM case.}
\end{figure}

The more interesting feature of such 
experiments is the possibility to give hints about
the viable SUSY--GUT scenarios, by constraining the
neutrino Yukawa sector. The reach of such
experiments as probes of different scenarios
are summarized in Table \ref{table_conc}
and displayed in Fig. \ref{ue3confronto}, where
we compare the scope of $\tau \to \mu \,\gamma$ and 
$\mu \to e \, \gamma$ experiments.

\begingroup
\squeezetable
\begin{table}[h]
\caption{\label{table_conc}Reach 
in $(m_0, m_{\tilde{g}})$  of the past, present and upcoming experiments from
their LFV sensitivity. LHC means that all the LHC testable 
parameter space will be probed; all means that soft masses up
to $(m_0, m_{\tilde{g}}) \lesssim 5$ TeV will be probed.}
\begin{ruledtabular}
\begin{tabular}{lcccc}
Experiment & \multicolumn{2}{c}{PMNS} &
\multicolumn{2}{c}{CKM}\\
&$t_\beta=40$& $t_\beta=10$ 
&$t_\beta=40$ &$t_\beta=10$
\\ \hline
{$\mu e$ sector}&&&&\\  
MEGA &
 \begin{tabular}{c} LHC \\  1.1 TeV\footnotemark[1]
 \end{tabular}& 
 \begin{tabular}{c}
 2 TeV   \\ no\footnotemark[1]
 \end{tabular}& no & no
\\ 
MEG &
  \begin{tabular}{c} all \\  LHC\footnotemark[1]
 \end{tabular}& 
 \begin{tabular}{c}
 all \\ 1.1 TeV\footnotemark[1] 
 \end{tabular}& 1.3 TeV   & no
\\ 
PRISM/PRIME\footnotemark[2]&
 all &  
 \begin{tabular}{c} all \\  LHC\footnotemark[1]
 \end{tabular}& 
 all  & 2.8 TeV
\\
$\tau\mu$ sector &&&&\\
BaBar, Belle & 
 1.2 TeV   & no & no & no
\\ 
SuperKEKB &
 2 TeV   & 0.9 TeV   & no & no
\\ 
Super Flavour\footnotemark[2]&
 2.8 TeV & 1.5 TeV   & 0.9 TeV  & no
\end{tabular}
\end{ruledtabular}
\footnotetext[1]{$U_{e3}=0$}
\footnotetext[2]{Post--LHC era, planned/discussed experiment}
\end{table}
\endgroup

Suppose that the LHC does find signals of low--energy supersymmetry, then
grand unification becomes a very appealing scenario, because of the successful 
unification of gauge couplings driven by the SUSY partners.
Among SUSY--GUT models, an $SO(10)$ 
framework is much favored as it is the `minimal' GUT to
host all the fermions in a single representations and
it accounts for the smallness of the observed neutrino
masses by naturally including the see--saw mechanism.
Moreover in the recent years $SO(10)$ SUSY models have
spurred much interest as in this framework it is possible
to build realistic fermion mass model and to account
for the proton lifetime bounds. In this paper we
have addressed the issue by a generic benchmark analysis,
within the ansatz that there is no fine--tuning in
the neutrino Yukawa sector.

From our analysis  we can state that lepton flavour
violation experiments should be able to tell us much about
the structure of such a SUSY--GUT scenario.
If they detect 
LFV processes, by their rate and exploiting the interplay
between different experiments, we would be
able to get hints of the structure of the
unknown neutrinos' Yukawas. In this sense, 
the capability
of a Super Flavour factory to discriminate between the minimal
mixing case and the $U_{e3}=0$ PMNS case
is a most interesting feature.

On the contrary, in the case that both MEG and a future Super Flavour
factory happen not to see any LFV process, only two possibilities
should be left: (i) a the minimal mixing, low $\tan\beta$ 
scenario; (ii) mSUGRA--$SO(10)$ see--saw without fine--tuned 
$Y_\nu$ couplings is 
not a viable
framework of physics beyond the standard model. Moreover, if the
planned, high sensitivity PRISM/PRIME conversion experiment, able to
test even the minimal mixing low $\tan\beta$ region, doesn't manage
to find LFV evidences, the latter conclusion should be the most sensible one 
and there should be no room left for the no fine--tuning 
framework we studied in this paper.
Actually one should remark that LFV experiments will be able
to falsify some of the SUSY GUT scenarios even in regions
of the SUGRA parameter space that are beyond the reach of
LHC experiments.
In this sense, the power of LFV experiments of testing/discriminating among
different SUSY GUTs models results very interesting
and highly complementary to the direct searches at the LHC.

\textbf{Acknowledgements}
SKV acknowledges support from Indo-French Centre for Promotion of
Advanced Research (CEFIPRA) project No:  2904-2 `Brane World
Phenomenology'.  He is also partially supported by INTAS grant,
03-51-6346, CNRS PICS \# 2530, RTN contract MRTN-CT-2004-005104
and by a European Union Excellence Grant, MEXT-CT-2003-509661. 
LC, AF and AM thank the PRIN `Astroparticle Physics' of the Italian
Ministry MIUR and the INFN `Astroparticle Physics' special project.
We also aknowledge support from RTN european program MRTN-CT-2004-503369
 `The Quest for Unification'.
LC thanks the Ecole Polytechnique--CPHT for hospitality. 
LC, AM and SKV also thank the CERN Theory Group for hospitality
during various stages of this work. We thank U. Chattopadhyay, S. Kraml,
S. Profumo, D. P. Roy and F. Zwirner for discussions. 
SKV also thanks T. Gherghetta for a reference.

\appendix*
\section{Notation and RGE equations}

\subsection{The model}

The model consists in a supersymmetric $SO(10)$ framework with the following 
breaking pattern
\begin{equation} 
SO(10) \stackrel{M_X}{\longrightarrow} SU(5)_{\mathrm{RN}} 
\stackrel{M_{GUT}}{\longrightarrow} \mathrm{MSSM}_{\mathrm{RN}}\nonumber
\end{equation} 
where $SO(10)$ is broken at the scale $M_X = 5\cdot 10^{17}$ GeV
which we equate it to the SUSY breaking mediation scale and the GUT scale is 
$M_{GUT} = 2 \cdot 10^{16}$ GeV. Below the scale $M_{X}$ the model is given by
the following $SU(5)_{\mathrm{RN}}$ superpotential
\begin{eqnarray}
W_{\mathrm{SU(5)_{\mathrm{RN}}}}& =& 
Y_{10\;ij}\; 10_{i} 10_{j} 5_{H} +  Y_{5\;ij}\; 
 10_{i}\bar{5}_{j} \bar{5}_{H}    \\ 
 && + Y_{1\;ij}\;  \bar{5}_{i} 1_{j} 5_{H} +
 M_{ij}\;  1_{i}1_{j} + \mu \bar{5}_{H} 5_{H}  \nonumber
\end{eqnarray}
While the corresponding soft SUSY breaking potential is
\begin{eqnarray}
V_{\mathrm{SU(5)_{\mathrm{RN}}}}&=& \left( A_{10\;ij} 10_i 10_j 5_H +
 A_{5\;ij}10_i \bar{5}_j \bar{5}_{\bar{H}}  \right. \nonumber \\ 
  && \left. + A_{1\;ij}\bar{5}_i1_j5_H +\tilde{M}_{ij}1_i 1_j 
  + B\mu 5_H \bar{5}_{\bar{H}} + \mathrm{h.c.}\right)\nonumber \\
 && +m^2_{\tilde{\bar{5}} \: ij} \bar{5}^*_i \bar{5}_j
 +m^2_{\tilde{10} \: ij} 10^*_i 10_j
 +m^2_{\tilde{1} \: ij} 1^*_i 1_j   \nonumber \\ && 
 +m^2_H 5^*_H 5_H
 +m^2_{\bar{H}} \bar{5}^*_{\bar{H}} \bar{5}_{\bar{H}}
 +M_5 \tilde{24}\tilde{24}
\end{eqnarray}

After reaching the GUT
scale, the theory is broken to the MSSM (plus right handed neutrinos) 
lagrangian
\begin{eqnarray}
 W_{\mathrm{MSSM_{RN}}}& =& Y^{u}_{ij}\; Q_{i} U^c_{j} H_2 +  
 Y^{d}_{ij}\; Q_{i} D^c_{j} H_1 +  Y^{e}_{ij}\;  L_{i} E^c_{j} H_1 
 \nonumber \\ &&  + Y^{\nu}_{ij}\;  L_{i}
 \nu^c_{j} H_2  + \mu H_1 H_2\\
V_{\mathrm{MSSM_{RN}}}&=& \left( A^{u}_{ij} \tilde{Q}_i \tilde{U}^c_j H_2 +
 A^{d}_{ij} \tilde{Q}_i \tilde{D}^c H_1 +    
  A^{\nu}_{ij} \tilde{L}_i \tilde{N}^c H_2 \right. \nonumber \\ &&
\left.  +\tilde{M}_{ij} \tilde{N}_i \tilde{N}_j + B\mu H_1 H_2 +
        \mathrm{h.c.}\right) \nonumber \\ &&
+m^2_{\tilde{Q} \: ij} \tilde{Q}^*_i \tilde{Q}_j 
 +m^2_{\tilde{U} \: ij} \tilde{U}^*_i \tilde{U}_j
 +m^2_{\tilde{D} \: ij} \tilde{D}^*_i \tilde{D}_j \nonumber \\ &&
 +m^2_{\tilde{L} \: ij} \tilde{L}^*_i \tilde{L}_j  
 +m^2_{\tilde{E} \: ij} \tilde{E}^*_i \tilde{E}_j  
+m^2_{\tilde{N} \: ij} \tilde{N}^*_i \tilde{N}_j \nonumber \\ &&
+m^2_H H^*_1 H_1  + m^2_H H^*_2 H_2 \nonumber \\ &&+ 
M_1 \tilde{B}\tilde{B}+M_2 \tilde{W}\tilde{W} +M_3 \tilde{G}\tilde{G}
\end{eqnarray}
The matching between $SU(5)$ parameters and MSSM ones, at $M_{GUT}$ 
is given by: 
\begin{eqnarray}
Y^{\nu}_{ij}  =  Y^{1}_{ij} & \; & Y^{u}_{ij} = 4 \: Y^{10}_{ij} \nonumber \\
Y^{d}_{ij}  =  \frac{1}{\sqrt{2}}  \: Y^{5}_{ij} &\;& Y^{e}_{ij} = 
\frac{1}{\sqrt{2}} \: Y^{5}_{ji} 
\end{eqnarray}
The matching of the soft A-matrices is the same as the Yukawas, whereas 
for the soft mass matrices is:
\begin{eqnarray}
m^2_{\tilde{U}} = m^2_{\tilde{10}}  \; ;  \; &
m^2_{\tilde{Q}} = m^2_{\tilde{10}}  \; ;  \; &
m^2_{\tilde{D}} = m^2_{\tilde{5}} \nonumber\\
m^2_{\tilde{L}} = m^2_{\tilde{5}}   \; ;  \; &
m^2_{\tilde{E}} = m^2_{\tilde{10}}  \; ;  \; &
m^2_{\tilde{N}} = m^2_{\tilde{1}}  
\end{eqnarray}
and
\begin{eqnarray}
M_1 = M_2 = M_3 = M_5
\end{eqnarray}

\begin{widetext}
\subsection{$SU(5)_{\mathrm{RN}}$ RGE}

{Conventions:} 
\begin{eqnarray}
\tilde{Y}=\frac{Y}{4\pi}; \;\;
\tilde{A}=\frac{A}{4\pi}; \;\;
\tilde{\alpha}=\frac{\alpha}{4\pi}=\frac{g^{2}}{\left(4\pi\right)^{2}};  \;\;
t=\ln\frac{M_{X}^{2}}{Q^{2}} \nonumber 
\end{eqnarray}
\\
{Yukawas:} 

\begin{eqnarray}
\frac{d}{dt}\tilde{Y}_{10\;ij} & = &  \frac{48}{5}\tilde{\alpha}_{5}
\tilde{Y}_{10\;ij} -24\; Tr\left(\tilde{Y}_{10}^{\dagger}\tilde{Y}_{10}\right)
\tilde{Y}_{10\;ij}-\frac{1}{2}Tr\left(\tilde{Y}_{1}^{\dagger}
\tilde{Y}_{1}\right)\tilde{Y}_{10\;ij} 
 -48\left(\tilde{Y}_{10}\tilde{Y}_{10}^{\dagger}\tilde{Y}_{10}\right)_{ij}
\nonumber \\
&&-\frac{1}{2}\left[\left(\tilde{Y}_{5}\tilde{Y}_{5}^{\dagger}
\tilde{Y}_{10}\right)_{ij}+\left(\tilde{Y}_{10}\tilde{Y}_{5}^{*}
\tilde{Y}_{5}^{T}\right)_{ij}\right]\\
\nonumber \\ 
\frac{d}{dt}\tilde{Y}_{5\;ij} & = &  \frac{42}{5}\tilde{\alpha}_{5}
\tilde{Y}_{5\;ij} - Tr\left(\tilde{Y}_{5}^{\dagger}\tilde{Y}_{5}\right)
\tilde{Y}_{5\;ij}-\frac{3}{2}\left(\tilde{Y}_{5}\tilde{Y}_{5}^{\dagger}
\tilde{Y}_{5}\right)_{ij} -24\left(\tilde{Y}_{10}\tilde{Y}_{10}^{\dagger}
\tilde{Y}_{5}\right)_{ij} \nonumber \\&&-\frac{1}{2}\left(\tilde{Y}_{5}
\tilde{Y}_{1}^{*}\tilde{Y}_{1}^{T}\right)_{ij} \\
\nonumber \\
\frac{d}{dt}\tilde{Y}_{1\;ij} & = &  \frac{24}{5}\tilde{\alpha}_{5}
\tilde{Y}_{1\;ij} - \frac{1}{2}Tr\left(\tilde{Y}_{1}^{\dagger}
\tilde{Y}_{1}\right)\tilde{Y}_{1\;ij} - 24\; 
Tr\left(\tilde{Y}_{10}^{\dagger}\tilde{Y}_{10}\right)\tilde{Y}_{1\;ij}
-3\left(\tilde{Y}_{1}\tilde{Y}_{1}^{\dagger}\tilde{Y}_{1}\right)_{ij}
\nonumber \\
&&-\left(\tilde{Y}_{5}^{T}\tilde{Y}_{5}^{*}\tilde{Y}_{1}\right)_{ij} 
\end{eqnarray}
\\
{Majorana mass:}

\begin{eqnarray}
\frac{d}{dt}M_{ij} =  -\frac{5}{2}\left[\left(M\tilde{Y}_{1}^{\dagger}
\tilde{Y}_{1}\right)_{ij}+\left(M\tilde{Y}_{1}^{T}
\tilde{Y}_{1}^{*}\right)_{ij}\right]
\end{eqnarray} 
\\
{Soft masses:} 

\begin{eqnarray}
\frac{d}{dt}\left(m^{2}_{\tilde{\bar{5}}}\right)_{ij} & = & 
\frac{48}{5}\tilde{\alpha}_{5}M^{2}_{5}\delta_{ij}- 
\left[ \left(m^{2}_{\tilde{\bar{5}}} \tilde{Y}_{5}^{\dagger}
\tilde{Y}_{5}\right)_{ij}+ \left(\tilde{Y}_{5}^{\dagger}\tilde{Y}_{5} 
m^{2}_{\tilde{\bar{5}}}\right)_{ij}\right]-\frac{1}{2}\left[ 
\left(m^{2}_{\tilde{\bar{5}}} \tilde{Y}_{1}^{*}\tilde{Y}_{1}^{T}\right)_{ij}
+ \left(\tilde{Y}_{1}^{*}\tilde{Y}_{1}^{T} 
m^{2}_{\tilde{\bar{5}}}\right)_{ij}\right]  \nonumber \\
&&-2 \left[ \left( \tilde{Y}_{5}^{\dagger} m^{2\;T}_{\tilde{10}} 
\tilde{Y}_{5}\right)_{ij}+\left(\tilde{Y}_{5}^{\dagger}\tilde{Y}_{5}\right)_{ij}
m^{2}_{\bar{H}}+\left(\tilde{A}_{5}^{\dagger}\tilde{A}_{5}\right)_{ij}\right] 
\nonumber \\ &&-\left[ \left( \tilde{Y}_{1}^{*} m^{2\;T}_{\tilde{1}} 
\tilde{Y}_{1}^{T}\right)_{ij}+\left(\tilde{Y}_{1}^{*}
\tilde{Y}_{1}^{T}\right)_{ij} m^{2}_{H} +\left(\tilde{A}_{1}^{*}
\tilde{A}_{1}^{T}\right)_{ij}\right] \\
\nonumber \\
\frac{d}{dt}\left(m^{2}_{\tilde{10}}\right)_{ij} & = & 
\frac{72}{5}\tilde{\alpha}_{5}M^{2}_{5}\delta_{ij}-24\left[ 
\left(m^{2}_{\tilde{10}} \tilde{Y}_{10}^{*}\tilde{Y}_{10}^{T}\right)_{ij}+ 
\left(\tilde{Y}_{10}^{*}\tilde{Y}_{10}^{T} m^{2}_{\tilde{10}}\right)_{ij}\right]
-\frac{1}{2}\left[ \left(m^{2}_{\tilde{10}} \tilde{Y}_{5}^{*}
\tilde{Y}_{5}^{T}\right)_{ij}+ \left(\tilde{Y}_{5}^{*}\tilde{Y}_{5}^{T} 
m^{2}_{\tilde{10}}\right)_{ij}\right] \nonumber  \\
&& -48\left[ \left( \tilde{Y}_{10}^{*} m^{2\;T}_{\tilde{10}} 
\tilde{Y}_{10}^{T}\right)_{ij}+\left(\tilde{Y}_{10}^{*}
\tilde{Y}_{10}^{T}\right)_{ij} m^{2}_{H} +\left(\tilde{A}_{10}^{*}
\tilde{A}_{10}^{T}\right)_{ij}\right] \nonumber \\
&&-\left[ \left( \tilde{Y}_{5}^{*} m^{2\;T}_{\tilde{\bar{5}}} 
\tilde{Y}_{5}^{T}\right)_{ij}+\left(\tilde{Y}_{5}^{*}
\tilde{Y}_{5}^{T}\right)_{ij} m^{2}_{\bar{H}} +\left(\tilde{A}_{5}^{*}
\tilde{A}_{5}^{T}\right)_{ij}\right]\\ 
\nonumber \\
\frac{d}{dt}\left(m^{2}_{\tilde{1}}\right)_{ij} & = & -\frac{5}{2}\left[ 
\left(m^{2}_{\tilde{1}} \tilde{Y}_{1}^{\dagger}\tilde{Y}_{1}\right)_{ij}+ 
\left(\tilde{Y}_{1}^{\dagger}\tilde{Y}_{1} 
m^{2}_{\tilde{1}}\right)_{ij}\right]
-5\left[ \left( \tilde{Y}_{1}^{\dagger} m^{2}_{\tilde{\bar{5}}} 
\tilde{Y}_{1}\right)_{ij}+\left(\tilde{Y}_{1}^{\dagger}
\tilde{Y}_{1}\right)_{ij} m^{2}_{H} +\left(\tilde{A}_{1}^{\dagger}
\tilde{A}_{1}\right)_{ij}\right] \\
\nonumber \\
\frac{d}{dt}\left(m^{2}_{H}\right) & = & \frac{48}{5}\tilde{\alpha}_{5}
M^{2}_{5}\delta_{ij}- 48 \left[Tr\left(\tilde{Y}_{10}^{\dagger}
\tilde{Y}_{10}\right)m^{2}_{H}+ 2 Tr\left(\tilde{Y}_{10}
m^{2}_{\tilde{10}}\tilde{Y}_{10}^{\dagger}\right)+
Tr\left(\tilde{A}_{10}^{\dagger}\tilde{A}_{10}\right)\right] \nonumber\\
&&-\left[Tr\left(\tilde{Y}_{1}^{\dagger}\tilde{Y}_{1}\right)m^{2}_{H}
+ Tr\left(\tilde{Y}_{1}^{\dagger}m^{2\;T}_{\tilde{\bar{5}}}\tilde{Y}_{1}
\right)+ Tr\left(\tilde{Y}_{1}m^{2}_{\tilde{1}}\tilde{Y}_{1}^{\dagger}
\right)+Tr\left(\tilde{A}_{1}^{\dagger}\tilde{A}_{1}\right) \right]\\
\nonumber\\
\frac{d}{dt}\left(m^{2}_{\bar{H}}\right) & = & \frac{48}{5}\tilde{\alpha}_{5}
M^{2}_{5}\delta_{ij}- 2 \left[Tr\left(\tilde{Y}_{5}^{\dagger}\tilde{Y}_{5}
\right)m^{2}_{\bar{H}}+ Tr\left(\tilde{Y}_{5}m^{2}_{\tilde{\bar{5}}}
\tilde{Y}_{5}^{\dagger}\right)+Tr\left(\tilde{Y}_{5}^{\dagger}
m^{2\;T}_{\tilde{10}}\tilde{Y}_{5}\right)+
Tr\left(\tilde{A}_{5}^{\dagger}\tilde{A}_{5}\right) \right]
\end{eqnarray} 
\\
{A-terms:} 

\begin{eqnarray}
\frac{d}{dt}\tilde{A}_{10\;ij} & = &  \frac{48}{5}\tilde{\alpha}_{5}
\left(\tilde{A}_{10\;ij}-2M_{5}\tilde{Y}_{10\;ij}\right) 
-24\;Tr\left(\tilde{Y}_{10}^{\dagger}\tilde{Y}_{10}\right)\tilde{A}_{10\;ij}
-\frac{1}{2}Tr\left(\tilde{Y}_{1}^{\dagger}\tilde{Y}_{1}\right)
\tilde{A}_{10\;ij}\nonumber\\&&
 -48\;Tr\left(\tilde{Y}_{10}^{\dagger}\tilde{A}_{10}\right)\tilde{Y}_{10\;ij} 
-Tr\left(\tilde{Y}_{1}^{\dagger}\tilde{A}_{1}\right)
\tilde{Y}_{10\;ij}-72\left[\left(\tilde{Y}_{10}\tilde{Y}_{10}^{\dagger}
\tilde{A}_{10}\right)_{ij}+\left(\tilde{A}_{10}\tilde{Y}_{10}^{\dagger}
\tilde{Y}_{10}\right)_{ij}\right]\nonumber \\&&-\frac{1}{2}\left[\left(
\tilde{Y}_{5} \tilde{Y}_{5}^{\dagger}\tilde{A}_{10}\right)_{ij}+
\left(\tilde{A}_{10} \tilde{Y}_{5}^{*}\tilde{Y}_{5}^{T}\right)_{ij}\right]
-\left(\tilde{Y}_{10} \tilde{Y}_{5}^{*}\tilde{A}_{5}^{T}\right)_{ij}-
\left(\tilde{A}_{5}\tilde{Y}_{5}^{\dagger}\tilde{Y}_{10}\right)_{ij}\\
\nonumber \\
\frac{d}{dt}\tilde{A}_{5\;ij} & = &  \frac{42}{5}\tilde{\alpha}_{5}\left(
\tilde{A}_{5\;ij}-2M_{5}\tilde{Y}_{5\;ij}\right) - Tr\left(
\tilde{Y}_{5}^{\dagger}\tilde{Y}_{5}\right)\tilde{A}_{5\;ij}-
2\;Tr\left(\tilde{Y}_{5}^{\dagger}\tilde{A}_{5}\right)\tilde{Y}_{5\;ij}\nonumber
\\&&-\frac{5}{2}\left(\tilde{Y}_{5}\tilde{Y}_{5}^{\dagger}
\tilde{A}_{5}\right)_{ij}-2\left(\tilde{A}_{5}\tilde{Y}_{5}^{\dagger}
\tilde{Y}_{5}\right)_{ij}-24\left(\tilde{Y}_{10}\tilde{Y}_{10}^{\dagger}
\tilde{A}_{5}\right)_{ij}-\frac{1}{2}\left(\tilde{A}_{5}
\tilde{Y}_{1}^{*}\tilde{Y}_{1}^{T}\right)_{ij}\nonumber \\
&&-\left(\tilde{Y}_{5}\tilde{Y}_{1}^{*}\tilde{A}_{1}^{T}\right)_{ij}
-48\left(\tilde{A}_{10}\tilde{Y}_{10}^{*}\tilde{Y}_{5}\right)_{ij} \\
\nonumber \\
\frac{d}{dt}\tilde{A}_{1\;ij} & = &  \frac{24}{5}\tilde{\alpha}_{5}
\left(\tilde{A}_{1\;ij}-2M_{5}\tilde{Y}_{1\;ij}\right) -\frac{1}{2}\; 
Tr\left(\tilde{Y}_{1}^{\dagger}\tilde{Y}_{1}\right)\tilde{A}_{1\;ij}
-24\;Tr\left(\tilde{Y}_{10}^{\dagger}\tilde{Y}_{10}\right)\tilde{A}_{1\;ij}
\nonumber \\ && -48\;Tr\left(\tilde{Y}_{10}^{\dagger}\tilde{A}_{10}\right)
\tilde{Y}_{1\;ij}-Tr\left(\tilde{Y}_{1}^{\dagger}\tilde{A}_{1}\right)
\tilde{Y}_{1\;ij}-\frac{11}{2}\left(\tilde{Y}_{1}\tilde{Y}_{1}^{\dagger}
\tilde{A}_{1}\right)_{ij}-\frac{7}{2}\left(\tilde{A}_{1}
\tilde{Y}_{1}^{\dagger}\tilde{Y}_{1}\right)_{ij}\nonumber \\
&&-2\left(\tilde{A}_{5}^{T}\tilde{Y}_{5}^{*}\tilde{Y}_{1}\right)_{ij}
-\left(\tilde{Y}_{5}^{T}\tilde{Y}_{5}^{*}\tilde{A}_{1}\right)_{ij} 
\end{eqnarray} 
\\
{$\mu$ terms:} 

\begin{eqnarray}
\frac{d}{dt}\mu^2 & = & 2 \;\left[\frac{24}{5}\tilde{\alpha}_5
 -12\: \textrm{Tr}\left( \tilde{Y}_{10}\tilde{Y}_{10}^\dagger \right) 
 -\frac{1}{2} \textrm{Tr}\left( \tilde{Y}_1\tilde{Y}_1^\dagger \right)
 - \textrm{Tr}\left( \tilde{Y}_5\tilde{Y}_5^\dagger \right) \right]\mu^2 \\
\nonumber \\
\frac{d}{dt}B\mu & = & - \left[\frac{48}{5}\tilde{\alpha}_5 M_5
 +12\: \textrm{Tr}\left( \tilde{A}_{10}\tilde{Y}_{10}^\dagger \right) 
 +\frac{1}{2} \textrm{Tr}\left( \tilde{A}_1\tilde{Y}_1^\dagger \right)
 + \textrm{Tr}\left( \tilde{A}_5\tilde{Y}_5^\dagger \right) \right]\mu 
\nonumber\\ && + \left[\frac{24}{5}\tilde{\alpha}_5
 -12\: \textrm{Tr}\left( \tilde{Y}_{10}\tilde{Y}_{10}^\dagger \right) 
 -\frac{1}{2} \textrm{Tr}\left( \tilde{Y}_1\tilde{Y}_1^\dagger \right)
 - \textrm{Tr}\left( \tilde{Y}_5\tilde{Y}_5^\dagger \right) \right]B\mu
\end{eqnarray}

\end{widetext}

\end{document}